\documentclass[aps, prd, twocolumn, superscriptaddress, nofootinbib, showpacs]{revtex4}
\usepackage{graphicx}
\usepackage{dcolumn}
\usepackage{amssymb}
\usepackage{amsmath}
\usepackage{array}
\usepackage{float}

\begin{document}

\newcommand{\Eq}[1]{\mbox{Eq. (\ref{eqn:#1})}}
\newcommand{\Fig}[1]{\mbox{Fig. \ref{fig:#1}}}
\newcommand{\Sec}[1]{\mbox{Sec. \ref{sec:#1}}}

\def\d{\mathrm{d}}

\def\mupn{^\mu_{\, \nu}}
\def\nn{\nonumber}

\def\K{\mathcal{K}}

\def\({\left(}
\def\){\right)}
\def\ie{{\it i.e. }}
\def\ms{M_6^4{}}
\def\mf{M_5^3{}}
\def\mq{M_4^2{}}
\def\pp{\partial_\mu\partial_\nu \pi}
\def\dbi{1+(\pa \pi)^2}
\def\tg{\tilde\gamma}

\newcommand{\detgt}{\sqrt{\tilde{g}}}
\newcommand{\detg}{\sqrt{g}}
\newcommand{\detgg}{\sqrt{\frac{\tilde{g}}{g}}}
\newcommand{\nabt}{\tilde{\nabla}}
\newcommand{\pa}{\partial}
\newcommand{\tmnt}{\tilde{T}^{\mu\nu}}
\newcommand{\tmn}{T^{\mu\nu}}
\newcommand{\mett}{\tilde{g}_{\mu\nu}}
\newcommand{\met}{g_{\mu\nu}}
\newcommand{\cS}{{\cal{S}}}
\newcommand{\cL}{{\cal{L}}}
\newcommand{\cE}{{\cal{E}}}
\newcommand{\HRule}{\rule{\linewidth}{0.4mm}}
\newcommand{\comment}[1]{}
\newcommand{\lagm}{{\cal{L}}_m}
\newcommand{\lage}{{\cal{L}}_E}

\def\O{{\cal O}}
\def\L{{\cal L}}
\def\mpl{M_{Pl}}

\newcommand{\ba}{\begin{eqnarray}}
\newcommand{\ea}{\end{eqnarray}}
\newcommand{\mn}{{\mu\nu}}
\newcommand{\cA}{{\cal A}}
\newcommand{\dof}{{\it dof} }
\newcommand{\eft}{{\it EFT} }

\newcommand{\eref}{\eqref}

\newcommand{\PHI}{\phi}
\newcommand{\PhiN}{\Phi^{\mathrm{N}}}
\newcommand{\vect}[1]{\mathbf{#1}}
\newcommand{\Del}{\nabla}
\newcommand{\unit}[1]{\;\mathrm{#1}}
\newcommand{\x}{\vect{x}}
\newcommand{\ScS}{\scriptstyle}
\newcommand{\ScScS}{\scriptscriptstyle}
\newcommand{\xplus}[1]{\vect{x}\!\ScScS{+}\!\ScS\vect{#1}}
\newcommand{\xminus}[1]{\vect{x}\!\ScScS{-}\!\ScS\vect{#1}}
\newcommand{\diff}{\mathrm{d}}

\newcommand{\be}{\begin{equation}}
\newcommand{\ee}{\end{equation}}
\newcommand{\bea}{\begin{eqnarray}}
\newcommand{\eea}{\end{eqnarray}}
\newcommand{\vu}{{\mathbf u}}
\newcommand{\ve}{{\mathbf e}}

        \newcommand{\vU}{{\mathbf U}}
        \newcommand{\vN}{{\mathbf N}}
        \newcommand{\vB}{{\mathbf B}}
        \newcommand{\vF}{{\mathbf F}}
        \newcommand{\vD}{{\mathbf D}}
        \newcommand{\vg}{{\mathbf g}}
        \newcommand{\va}{{\mathbf a}}


\title{Tests of Modified Gravity Theories in the Solar System}

\newcommand{\addressImperial}{Theoretical Physics, Blackett Laboratory, Imperial College, London, SW7 2BZ, United Kingdom}

\author{Ali Mozaffari}
\email{ali.mozaffari05@imperial.ac.uk}
\affiliation{\addressImperial}

\date{\today}

\begin{abstract}
We review the case for testing preferred acceleration scale theories of gravity
(sometimes falling under the guise of MOdified Newtonian Dynamics) in the Solar System using the forthcoming LISA Pathfinder (LPF) mission.  Using a combination of analytical and numerical results, we suggest that different types of theory should be detectable using the predicted anomalous tidal stresses effects around the saddle points of the Newtonian gravitational field.  The saddle point bubbles expected extent of $\sim 400$ km are to be contrasted with potential miss parameters of $\leq 10$ km, making such a test in easy reach of LPF.  We also consider routes to constraining our theories from data, based on scenarios of both null and positive results.   \end{abstract}


\maketitle

\section{Introduction}

The concordance model of modern cosmology rests soundly on two cornerstones, a universe filled mostly with cold dark matter (CDM) and dark energy (described
by a cosmological constant), i.e. $\Lambda$CDM, with underlying dynamics characterised by Einstein's theory of General Relativity (GR).  Whilst this model explains the early universe with ever increasing accuracy~\cite{Planck}, as long as there remains the lack of direct detection of a dark matter particle (baring unviable candidates such as neutrinos~\cite{DMastro,DMthesis}), it remains prudent to consider alternatives.  One such pathway available is to modify the underlying dynamics themselves, subject to the condition that above certain scales we restore our familiar Newtonian limit.  MOdified Newtonian Dynamics (MOND) provides just such a scheme.  The MONDian paradigm seeks to explain away galactic dynamics through the use of a modified force law, introducing
a preferred acceleration scale, on the scale of typical galactic accelerations
(see~\cite{MONDreview} for a detailed review).  On galactic scales, these modified effects become dominant, but at larger accelerations, gravity becomes idyllically described by Newtonian dynamics.

Although ideas of ``modifying'' gravity are in some way nothing new,
it was Milgrom in 1983 who first proposed a theory of modified inertia~\cite{Milgrom:1983ca}.
 This was subsequently developed in 1986 into the theory known as AQUAL~\cite{aqual} by Bekenstein and Milgrom, formulating a Lagrangian theory which would satisfy
energy and momentum conservation.  Investigating the equations of motion from that leads us to a modified Poisson relation - a common way to present such theories.  In the past decade, the potential accomplishments of MONDian theories have been put on an equal pedestal to GR with the development of fully relativistic modified gravity (MG) theories.   We find in the literature now a litany of examples of such\footnote{The original work on AQUAL~\cite{aqual} did
describe a relativistic extension for MOND, however it was soon realised
it could not take into account observations of light deflection from galaxies
nor could it properly restrict the tachyonic behaviour of its field.}, starting in 2004 with Bekenstein's ground breaking theory of T$e$V$e$S~\cite{teves}.
 T$e$V$e$S attempted to overcome previous issues in this field by introducing
 a vector and scalar field into the mix, fixing acausal and light deflection
 issues, at least at the payoff of having to fix more variables.  Similarly
 with the Lorentz violating work of Einstein \AE ther theories~\cite{jacobmatAE, aether}\footnote{In fact the original Einstein \AE ther theory reduces just to Newtonian dynamics in its weak field, but its construction introduces an acceleration scale, a feature that was later used in generalisations to reduce to MOND.}, these various ideas were expanded on and generalised by Zlosnik, Ferreira and Starkman in 2006~\cite{aether1,aether2,AESS}, as well
as attempts by Skordis and others to generalise and investigate the cosmology of these theories~\cite{kostasrev,SkordisGenTeves,tevesstrucform}.
Since 2009, Milgrom has produced bimetric theories~\cite{bimetric}, motivating
a quasi-linear MONDian theory from a relativistic perspective.  There have
been various other ideas for gravity theories in this way~\cite{BSTV,Clifton11,Fam-gaugh}.  Whilst the MONDian paradigm provides a useful framework for making connection to observables, the free functions and parameters in these theories remain relatively unconstrained, leading problems of fine tuning.  These theories
have the danger of explaining an observations at the cost of being purely
empirically fitted to data.  Much work has been done investigating these modified effects on the largest scales, for instance applying constraints from galactic data when seeking dark matter alternatives~\cite{Zhao,binney,yusaf,yusaf1,yusaf2}.
 The much hailed Bullet Cluster (1E 0657-558) has been considered for what it can tell us about the necessity or needlessness of dark matter and MOND~\cite{Angus,bullet,bullet1,bullet2,bullet3}.
 These gravitational lensing studies in the past decade have suggested that
 CDM fits the data very well and modified gravitational force laws are statistically unlikely to explain away the results.  There
remains however a lack of consensus on interpreting the weak lensing survey
and also there are clusters, such as Abell 520~\cite{AbellDM}, which are not easily explained by any current paradigm.  Quite a different tack has come from applying Lorentz violating mechanisms (typically well constrained in the matter sector) to the gravity sector~\cite{withers, lviolationcosmo}.  Constraints from high
energy experiments, such as those at the LHC, especially in the light of the most recent data, have provided some of the best detailed constraints to be seen in the Solar System.  Perhaps a good way to investigate general modified non-relativistic theories is to examine deviations from the inverse square law, as considered in~\cite{Blanchet, Sereno,Milgromss}.  Little more
however seems to be known about constraining modified gravity theories purely in the Solar System.  

A more pragmatic way to approach these issues is to consider that there appears to be a ubiquitous acceleration scale in the universe, $ a_0\sim 10^{-10}\unit{ms}^{-2}$.  It crops up variously in cosmology and astrophysics, e.g. the cosmic expansion rate and galactic rotation curves appear curiously linked to this value.
 Such an observation has prompted the investigation of alternative theories of gravity endowed with such a preferred acceleration, whatever its eventual
physical effect.  Such ideas were first proposed with the motivation of bypassing the need for dark matter, but they may also be considered independently from this, simply as mature alternative theories of gravity~\cite{Clifton11} into which this acceleration scale has been embedded.  In such a guise, they constitute prime targets for experimental gravitational tests inside the Solar System.  A chance of extending the forthcoming LISA Pathfinder (LPF) mission~\cite{bekmag,LISA,companion}, to include probing the low acceleration regime around gravitational saddle points (SP), appears to provide just such an opportunity, both for testing and also cleanly constraining these theories.
  
We organise this paper as follows, in Section \ref{techniques} we will consider on both analytical (Section \ref{Uformalism}) and numerical (Section \ref{SOR}) investigations in to preferred acceleration scale theories.  Section \ref{LISApathfinder} introduces the LISA Pathfinder spacecraft and Section \ref{GWtech} shows how methods from experimental gravitational wave searches can be applied to characterise such a test, with results for various theories in Section
\ref{secsnr}.  Section \ref{constrain} attempts to explore the wider parameter space of these theories, varying both constants as well of the free function itself, in order constrain such theories from data.  We conclude with some future thoughts and directions in this field.   

\subsection{Finding $a_0$ - Saddle Points in the Solar System}

Here we will introduce the techniques we will need later to characterise
theoretical and experimental ideas in MONDian tests.  We will follow the notation and formalism first developed in~\cite{bekmag}, as well as numerical
ideas presented in~\cite{bevis}.  Obviously to test MONDian theories, we will need a regime where the {\it total} acceleration on test masses will be small enough to be approaching galactic acceleration scales, which we will take as $a_0$.  Such regions do in fact exist in the Solar System, our own cosmic backyard.  Before we continue, we will need to understand where these regions are located and solve our MONDian equations of motion in these regimes, before examining how we can test these ideas concretely.  
 
We start by considering a two body gravitational system, with masses $m$
and $M$, such that $M \gg m$, separated by some distance $R$ along the $\ve_z$ axes linking them.  We centre the coordinates on mass $M$ and look at the resultant acceleration along $\ve_z$, \be \vF_N = -\Del \Phi_N = \left(-\frac{G M}{r^2} + \frac{G m}{(R-r)^2}\right)\ve_z \label{2bodyFN}\ee  The stationary point of this force is thus located at \be r_{s} = \frac{R}{1 + \sqrt{m/M}}\simeq R\left(1 - \sqrt{\frac{m}{M}}\right)\ee
 The form of the force shows that moving along $\ve_z$ towards either mass
results in an attractive force, however moving perpendicular to the axes results in a restoring force towards the stationary point - we have a gravitational saddle point (SP).  We should be clear to distinguish these points from the well known Lagrange points, which exist {\it only} in a system of rotating
bodies,
 whereas this saddle always exists (the effect of two attractive forces along
 the line linking them, in opposite directions).  We find that the Newtonian force is linearised about the saddle, taking
the form \be \vF_N = -\Del \Phi_N = A\,(r-r_{s})\,\ve_z \label{linearFN}\ee where $A$ is the Newtonian tidal stress at the saddle, defined as the derivative of the force \be S_{ij}^N = \frac{\partial^2 \Phi_N}{\partial x_i \partial x_j}\ee Here $S_{ij}^N $ is simply a constant, found when we compute the Taylor expansion coefficients in the linear expression (\ref{linearFN}) from the
full two body expression (\ref{2bodyFN}) \be A = 2\frac{G M}{r_s^3}\left(1
+ \sqrt{\frac{M}{m}}\right) \label{2bodyNewtS}\ee  We can make two observations, one being that since $F_N \rightarrow 0$, it will indeed pass through the acceleration barrier of $a_0$, suggesting MONDian effects should be visible around saddles.
 For the Earth-Sun SP, such a low acceleration region is located at $r \leq 2.2$m around the saddle - a poor prospect for a satellite target.  If however we consider the rule of thumb for MONDian systems, i.e. \bea F \leq a_0 &\Rightarrow& F\rightarrow\sqrt{F_N a_0}\eea then the (previously linear) force near the SP is now of the form \be F \rightarrow \sqrt{A a_0 |r-r_s|}\ee and the tidal stresses look like \be S = \frac{\partial F}{\partial _r} \sim \frac{1}{\sqrt{r
- r_s}}\ee it would appear these diverge as we approach the saddle!  Clearly we need to investigate the calculation using a fully relativistic theory, but this simple calculation provides at least a proof-of-concept for a tidal stress based MOND saddle test.  

A second relevant point to make concerns the other contributions to the Newtonian
tidal stresses at the saddle, surely the Solar System and the galaxy will play a role here?  At leading order, only the Earth and Sun play a role in this calculation, as (\ref{2bodyNewtS}) shows.
 The effect of the Moon, providing a truly 3-body system, can be computed
 using a numerical treatment of the saddle system, as we will shortly show
 in Section \ref{SOR}.  One conclusion of that work is that the position
 of the Earth-Sun saddle is shifted with respect to the phase of the Moon
 (and hence at different times of the month the saddle is shifted to a known, but differing location), on the order of a few tens of km.  Taking the effect of most of the mass of the solar  system (from Saturn and Jupiter) into account shifts the saddle a few more km.  Taking the contribution from the galaxy
into account shifts it a tiny bit more.  Given this, we can consider the
total Newtonian tidal stress at the saddle taking the form \be A_{SP} \simeq
A_{ES} + A_{M} + A_{SS} + A_{G} + \dots\ee where $ES$ denotes Earth-Sun,
$M$ denotes Moon, $SS$ denotes Solar System and $G$ denotes the galactic contribution.  The ordering here is such that each contribution is smaller in magnitude than the one previous.  Given that each contribution to the
saddle is an attractive force component, there will always be a saddle (and
at a location close to the 2-body case) and hence an observable for a tidal
stress experiment.

\subsection{Classifying MONDian theories}\label{theory}
In the wider modified gravity literature, one can find a large number of relativistic modified gravity theories.  Their complexity and differences arise from the requirement that they should explain relativistic phenomena (such as lensing and structure formation) without appealing to dark matter, whilst in the non-relativistic regime have some MONDian and Newtonian
limit.  In general, the large profusion of relativistic MONDian theories reduce to just three different non-relativistic limits:

Our job is to approach theories where such modified behaviour is present and see if they represent good prospects for detection.  Their complexity and differences arise from the requirement that they should explain relativistic phenomena (such as lensing and structure formation) without appealing to dark matter, whilst in the non-relativistic regime have some Newtonian and
other modified limit. The manner in which such effects are manifest may however vary widely and there have been many previous studies as to the phenomenology of these ideas, particularly in this non-relativistic regime~\cite{typeIIpaper,ali,Milgromss,aether,teves}.
 We will briefly outline some of these here, with the caveat that this list
 is neither exhaustive, nor represents the final story on gravity theories
 at the time of writing and for a more in depth look at gravity theories,
 we point the reader towards~\cite{Clifton11}.
 
 \begin{itemize}
\item{\bf Type I:} Here the total potential acting on non-relativistic particles
is given by the sum of the usual Newtonian potential $\Phi_N$ and a fifth force field, $\phi$: \be \Phi = \Xi\Phi_N+\phi\ee where $\Xi$ is some constant
usually set to unity and the Newtonian potential satisfies the usual Poisson equation \mbox{$\nabla^2 \Phi_N=4\pi G \rho$}, and the field $\phi$ is governed by: \be \nabla \cdot \left(\mu(z)\nabla \phi\right) = \kappa G \rho \label{type1} \ee The argument of $\mu(z)$ is given by \be z=\frac{\kappa}{4\pi}\frac{\vert\nabla\phi\vert}{a_0} \ee where $\kappa$ is a dimensionless coupling constant.  $\mu$ is a free function,
typically chosen limits of the theory are $\mu\rightarrow 1$ when $z\gg 1$ and $\mu \simeq z$ for $z\ll 1$.  The effect of these fields is twofold, in the large $z$ regime, $\phi \rightarrow \frac{\kappa}{4\pi} \Phi_N$ mimicking the Newtonian, this makes the physical potential have the form \be \Del \Phi \rightarrow \left(\Xi + \frac{\kappa}{4\pi}\right)\Del\Phi_N\ee
or equivalently the form of Newton's constant is altered \be G_{ren} \rightarrow
\left(\Xi + \frac{\kappa}{4\pi} \right)G_N \ee Cosmology sets bounds on the
variation of $G$, from BBN and effects in the CMB~\cite{CarrollLim, Nconstraint}.
 
 Additionally these two fields mean that the Newtonian behaviour is always
 present in the non-relativistic regime and non-linear behaviour in $\phi$ gets triggered at a certain acceleration \be a_{\text{trig}} = \left(\frac{4\pi}{\kappa}\right)^2
a_0\ee The field however remains sub-dominant until $a_N = a_0$ and this
is when fully modified behaviour is seen (in the galactic regime).  It is
this onset of non-linearity that we hope to probe with LPF.

\item{\bf Type II:} These are similar in set-up to type {\bf I}, with $\Phi = \Phi_N +\phi$ and
$\phi$ governed by a driven linear Poisson equation:\be \nabla^2 \phi = \frac{\kappa}{4\pi}\nabla\cdot\left(\nu(w)\nabla \Phi_N \right) \label{type2} \ee The argument of $\nu$ is given by \be w = \left(\frac{\kappa}{4\pi}\right)^2\frac{\vert\nabla\Phi_N \vert}{a_0}  \ee Once again $\nu$ is a free function and typically we give it the form
$\nu\simeq 1/\sqrt{w}$ for $w\ll 1$ and $\nu \rightarrow \unit{constant}$
for $w\gg 1$.  We divide this up in the subtypes of {\bf IIA} or{ \bf IIB} with qualitatively very different implications, which can we seen more clearly if we return to using the physical
potential form \bea \Del^2 \Phi &=& \Del \cdot\left( \hat{\nu} \Del\Phi_N\right)
\\  \hat{\nu} &=& 1 + \left(\frac{\kappa}{4\pi}\right)\nu\eea 
Consider in the large $w$ regime: \begin{itemize}
\item{In type {\bf IIA}, $\nu \rightarrow 0$ which implies no $G$ renormalisation occurs and $a_{\text{trig}} = a_0$.  The whole theory in fact hinges on $\Phi$, all other fields are considered auxiliary.}
\item{In type {\bf IIB}, $\nu \rightarrow 1$ means a trigger acceleration similar to type {\bf I}.  }\end{itemize}

\item{\bf Type III:} Crucially, here non-relativistic particles are sensitive to a single field $\Phi$, satisfying a non-linear Poisson equation: \be \nabla\cdot\left(\tilde{\mu}(x)
\nabla \Phi \right) = 4 \pi G \rho \label{type3} \ee where the argument of
$\tilde\mu$ is \be x = \frac{\vert\nabla\Phi\vert}{a_0}\ee so that $\tilde \mu\rightarrow 1$ when $x\gg 1$ and $\tilde \mu\sim x$ for
$x\ll 1$. Again no renormalisation of $G$ and a trigger acceleration
$a_{\text{trig}} = a_0$
\end{itemize}
As the trigger acceleration sets the scale of the SP bubble, using the current estimates for our parameters ($\kappa = 0.03$, $a_0
= 10^{-10}\unit{ms}^{-1}$) we find these to be \bea {\bf I,IIB}:\nonumber\\a_{\text{trig}} &=& \left(\frac{4 \pi}{\kappa}\right)^2 a_0 \simeq 10^{-5}\unit{ms}^{-2} \Rightarrow r_0\sim 383 \unit{km}\nonumber\\{\bf IIA,III}:\nonumber\\ a_{\text{trig}} &=& a_0 = 10^{-10} \unit{ms}^{-2} \Rightarrow r_0
\sim 2.2 \unit{m}\nonumber\eea
These distinctions group together types {\bf I} and {\bf IIB} as the best candidates for detection with LPF; types {\bf IIA} and {\bf III} would easily escape any negative result.

An important distinction here stems from the fact that
we have a curl term (often called a magnetic field) in type {\bf I} and {\bf
III} theories.  This is easiest seen when one attempts to linearize the non-linear Poisson equations present by introducing an auxiliary vector field (e.g. $\mu\nabla\phi$ for type {\bf I} theories) - such a field has non-zero curl. The same is not true for type {\bf II} theories, being already linear in $\phi$ and driven by a function of the Newtonian field, $\nu\nabla \Phi_N$, (a quantity which has a curl). This turns out to have a significant quantitative effect upon the magnitude of the saddle tidal stresses, as the magnetic field is known to soften the anomalous tidal stresses around the saddle points in type {\bf I} theories.

A scan of the relativistic MONDian theories proposed in the literature suggests
that they fall into these categories. Bekenstein's TeVeS~\cite{teves} as well as Sanders' stratified theory~\cite{BSTV} have type I limits. Milgrom's Bimetric theory~\cite{Milgrom:2009gv,Milgrom:2010cd} can be either type I or type II, depending on details. GEA theories~\cite{aether,aether1} and
Galileon k-mouflage~\cite{k-mouflage} have a non-relativistic limit of type III. Often authors have attended to different considerations and constraints, so the parameter $\kappa$ has been taken to be different. However, as we
will point out, if in each case the same considerations have been employed, the value of $\kappa$ would have to be comparable.  

\section{Analytical and Numerical Results}\label{techniques}
Next we move onto using these different types of theory to compute observes
at the SP.  First one can use analytical results to shed some light on the
predicted signals available, followed by a full numerical treatment and finally
considering how to feasibly measure such stress signals.  

\subsection{Type I - $\vect{U}$ Formalism}\label{Uformalism}
We move to a system of spherical polar coordinates, centered on the saddle.
 Clearly here $\Del^2\Phi_N = 0$, so we can consider a multipole expansion
for $\Del\Phi_N$, truncated at linear order: \bea -\Del \Phi_N &=& \vF_N = A\, r\, \vN \label{linearNewt}\\ \vN &=& N_r\ve_r + N_\psi \ve_\psi \\ N_r &=& \frac{1}{4}(1 + 3 \cos 2\psi ) \\ N_\psi &=& -\frac{3}{4} \sin 2\psi \label{linearNewt}\eea Notice due to the symmetries of this two body system, the polar angle $\varphi$ does not appear (but would be important if a three body system, such as including the Moon, was considered).

Recall the non-linear modified Poisson equation (\ref{type1}) for the MONDian field $\phi$, let's move to a linear system of variables by defining \be \vU = -\frac{\kappa}{4\pi}\frac{\Del\phi}{a_0}\mu
\label{U defn}\ee meaning $U = \mu z$.  Since our free functions here are $\mu = \mu(z)$, we can similarly write them solely as $\mu = \mu(U)$. This change of variable allows us to write dimensionless vacuum equations \bea \Del \cdot \vU &=& 0 \label{vect1} \\ 4\, m\, U^2 \Del \wedge \vU + \vU \wedge \Del U^2 &=& 0 \label{vect2}\eea where $4m$ has the form \be 4m = \frac{d \ln U^2}{d \ln \mu}\ee and we have dropped sources (as would be the case at the SP).  We can reconnect with the MONDian force from the expression \be -\Del\phi = \delta\vF = \frac{4 \pi a_0}{\kappa}\frac{\vU}{\mu(U)} \label{MONDforceexpansion}\ee 
\begin{figure}[t!]\begin{center} \resizebox{1\columnwidth}{!}{\includegraphics{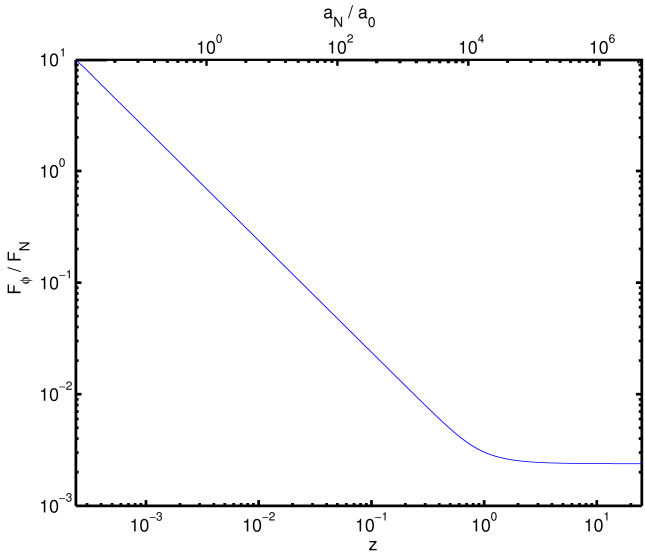}}
\caption{\label{fig:Fphi/FN plot analytical mu}{Log plot of ratio between the MONDian and Newtonian forces, $F_\phi/F_N$, against $z=(k / 4\pi) |F_\phi| / a_0$ (bottom axis) and $F_N/a_0$ (top axis) using a free function which
smoothly interpolates between the large and small $z$ regimes.  So that $F_N\sim F_\phi$ when $F_\phi\sim a_0$ (and so $z=\kappa/4\pi$; also $F_N\sim a_0$) and at the same time have $F_\phi/F_N\sim \kappa  /4\pi\ll 1$ in the Newtonian regime ($z\gg 1$, $F_N\rightarrow\infty$), we must trigger MONDian behaviour in $\phi$ at accelerations much larger than $a_0$ (when $z\sim 1$).}} \end{center}
\end{figure} 
For our choice of free function $\mu$, we will start with
the case considered previously in analytical~\cite{bekmag} and numerical work~\cite{bevis}, namely \bea z = \frac{\mu}{\sqrt{1 - \mu^4}} &\Longleftrightarrow&
\mu = \sqrt{\frac{\sqrt{1 + z^4}-1}{2z^2}} \nonumber\\ &\Longleftrightarrow&
\mu = \frac{U^{1/2}}{(1 + U^2)^{1/4}} \label{mufid}\eea It is clear from $U = \mu z$ and the behaviour of $\mu$ that each limit satisfies \bea z \gg 1 &\Rightarrow& U \gg 1\,\,\,\,\, \text{(Quasi-Newtonian)}\nonumber \\ z \ll 1 &\Rightarrow& U \ll 1 \,\,\,\,\, \text{(Deep-Mondian)}\nonumber\eea Between these two regions, we have a boundary located at $|\vU|^2 \simeq 1$.  To locate these regions, we assume \be \mu \rightarrow 1 \Rightarrow \Del\phi \simeq \frac{\kappa}{4\pi} \Del\Phi_N \ee which will clearly be true at zeroth order and as we will show later, leading order corrections are negligible at the boundary. Using expression (\ref{linearNewt}), this becomes \be |\vU|^2 = \left(\frac{\kappa}{4\pi} \frac{|\Del\phi|}{a_0}\right)^2 \simeq \left(\left(\frac{\kappa}{4\pi}\right)^2 \frac{|\Del\Phi_N|}{a_0}\right)^2 \simeq 1\ee this takes the form (after rearrangement) \be r^2 \left(\cos^2 \psi + \frac{1}{4} \sin^2 \psi\right)^2 = \left(\frac{16\pi^2}{\kappa^2}\frac{a_0}{A}\right)^2
= r_0^2\ee which is just the equation for an ellipsoid with a size we will
denote $r_0$.  These results show that in general the functional forms of the inner and outer ellipsoid (hereafter bubble) solutions should be quite different.

\subsubsection{Quasi-Newtonian (QN) Regime}\label{QN Analytic}
Given this system of vector equations, we need to specify boundary conditions.
 For $r/r_0 \gg 1$, we expect $\mu \rightarrow 1$ and so the MONDian potential to mimic the Newtonian $\phi \approx \frac{\kappa}{4\pi}\Phi_N$.  Let's pick our ansatz to be of the form \bea \vU &=& \vU_0 + \vU_2 \\ \vU_0 &=& \frac{r}{r_0}\vN(\psi) \eea and $\vU_2$ will be some subdominant contribution as we move far from the saddle, but a very relevant one closer to the bubble.  Additionally although $\vU_0$ is curl free, the form of Equation (\ref{vect2}) suggests $\vU_2$
will in general have a curl, automatically satisfying \bea \Del \cdot \vU_2 &=& 0 \label{divU2} \eea and from rearrangement of (\ref{vect2}) we see that at leading order, $\vU_2$ is sourced
by $\vU_0$
\bea \Del \wedge \vU_2 &=& -\frac{\vU_0 \wedge \Del|\vU_0|^2}{4m\,|\vU_0|^{2}} \label{curlU2} \eea  Using the notation \be \vU_2 = U_r\ve_r + U_\psi\ve_\psi \ee The form of (\ref{divU2}) and (\ref{curlU2}) strongly suggest that $U_2 \propto
1/r$ and so we can write \be \vU_2 = \frac{r_0}{r}\vB(\psi) = \frac{r_0}{r}\left(F(\psi)
\ve_r + G(\psi) \ve_\psi\right)\ee In this case we find \bea F &=& {2\over 5+3\cos2\psi} - \frac{\pi}{3\sqrt 3} \label{sol1}\\ G\sin\psi &=&{\tan^{-1}(\sqrt 3 -2 \tan{\psi\over 2})+
\tan^{-1}(\sqrt 3 +2 \tan{\psi\over 2})\over \sqrt 3}\nonumber \\ &-& \frac{\pi}{3\sqrt 3}\left(\cos\psi + 1\right) \eea where we have used the conditions
of homogeneity and continuity and that at the boundaries of the bubbles,
$G(\psi=0) = G(\psi=\pi)=0$ (akin to the Newtonian).  Expanding (\ref{MONDforceexpansion}) in this limit gives us  \bea \delta{\mathbf F} &=& -\nabla\phi = \frac{4 \pi a_0}{\kappa}\frac{\vU}{\mu}
\\\nonumber &\simeq& {4\pi a_0\over \kappa} \left(\underbrace{\frac{r}{r_0}{\mathbf N}}_{G_N\, renorm} + \underbrace{\frac{r_0}{r}\left({{\mathbf N}\over 4 N^2}+{\mathbf B}\right)}_{main \, observable} + \dots \right)\eea Additionally, we can justify our prior assumption of $\vU = \vU_0$ when estimating the bubble boundary.  Given $|\vB| \sim |\vN| \sim \mathcal{O}(1)$, our naive first order correction would be \be \frac{|\vU_2|}{|\vU_0|} \simeq \left(\frac{r_0}{r}\right)^2 \nonumber\ee however this is only strictly true
in the $r/r_0 \gg 1$ limit, so we must think more carefully about our assumption
at $r\simeq r_0$.  We assumed $\mu \rightarrow 1, \Del\phi \rightarrow \frac{\kappa}{4\pi}\Del\Phi_N$,
however in reality we have $\mu = 1 - \delta \mu$ in this limit and so the
fractional correction is of order \be \frac{F_\phi}{F_N} \simeq \frac{\kappa}{4\pi}
\left(1 + \frac{1}{4z^2} +\dots \right)\ee Expanding to first order gives \be \vF_\phi = \frac{\kappa}{4\pi}\vF_N + \vF_\phi^{(1)}\ee
and so \be \frac{\delta F}{F_N} \sim \frac{\kappa}{4\pi}\left(\frac{r_0}{r}\right)^2\ee
meaning that even close to the boundary, taking $\vU \sim \vU_0$
is a good approximation as long as $\kappa \ll 4\pi$ is true.  In circumstances
when this is not the case, our approximation will break down.  The bubble
however will also be much smaller (remember $r_0 \sim 1/\kappa^2$) and
so after a few $r_0$, we will clearly be in the QN regime anyway.  For the
Solar System at least, it is doubtful that assuming such will result in an order of magnitude correction in $r_0$.

\subsubsection{Deep-MONDian (DM) Regime}

Our previous intuition with boundary conditions does not help us here, since
we expect a very different signal compared to the linear Newtonian falling
to zero at the saddle.  We see that if $4m \rightarrow$ constant, then (\ref{vect1}) and (\ref{vect2}) have scaling symmetries \bea \vU \rightarrow \vU \nonumber\\ r \rightarrow \lambda\, r \eea which suggests an ansatz for the potential as \be \vU = C \left(\frac{r}{r_0}\right)^{\alpha-2} (F(\psi) \ve_r + G(\psi) \ve_\psi) \ee where $\alpha - 2$ is used for notational convenience later
and $C$ is a constant required for matching between the two regimes.  We
will look for solutions which keep $U$ small but have tidal stresses become increasingly divergent as $r/r_0 \ll 1$. This results in a solution \be \delta \vF \approx \frac{4\pi a_0}{\kappa}\, C^{1/2}\left(\frac{r} {r_0}\right)^{\frac{\alpha-2}{2}} \frac{\vect{D}(\psi)}{D^{1/2}}
\label{gradphi DM rD}\ee where $\vect{D}$ is the angular profile here.  Also requiring $U \ll 1$ means $\alpha > 2$ in all cases (a point realised but not explicitly spelt out in~\cite{bekmag}), whilst $\alpha < 4$ is needed for a divergent tidal stress solution.  Since we have no restriction on our scaling $C$, without loss of generality we set $F(\psi = 0) = F(\psi = \pi) = 1$ and again enforcing the condition $G(\psi = 0) = G(\psi = \pi) = 0$ gives us solutions $\alpha \approx 3.528$, with profile functions \bea F &\approx& 0.2442 + 0.7246 \cos 2\psi + 0.0472 \cos 4\psi + \dots \nonumber \\ G &\approx& -0.8334 \sin 2\psi - 0.0368 \sin 4\psi + \dots \eea

\subsection{Type II - Laplacian}\label{typeIIsection}
  We will only consider type IIB theories here, the IIA case has been considered separately~\cite{typeIIpaper}, which suggested such theories represent poor targets for an LPF test.  In these theories, we have the driven Poisson equation \be \Del^2 \phi = \frac{\kappa}{4\pi}\Del\cdot \left(\nu(w)\Del\Phi_N\right)\nonumber\ee
with argument \be \nonumber w = \left(\frac{\kappa}{4\pi}\right)^2 \frac{|\Del\Phi_N|}{a_0}\ee
such that $\nu \rightarrow 1/\sqrt{w}$ for $ w \ll 1$ and $\nu \rightarrow
1$ for $w \gg 1$.  One noteworthy point here is that we are faced with a driven Poisson equation with a known and well understood right hand side,
hence computing solutions here are far easier than with the non-linear type I equation.  Whilst some effects are more precisely model dependent than others, we suggest the function \be \nu = \left( 1 + \frac{1}{w^2}\right)^{1/4}\ee
because it draws similarities to the type I free function of (\ref{mufid}).
 Given Equation (\ref{type2}), let's expand \be \Del^2 \phi = \Del\cdot(\nu\Del\Phi_N)
= \nu \underbrace{\Del^2 \Phi_N}_{= 0|_{SP}} +\, \Del \nu \cdot \Del \Phi_N\ee
and then use the linear Newtonian approximation, providing the form of the source term  \bea\Del^2\phi = \frac{a_0}{2}\left(\frac{4\pi}{k}\right)\left(\frac{1}{r_0 \,r}\right)^{1/2}\left(\frac{r_0^2}{r_0^2 + (rN)^2}\right)^{1/4}\nonumber
\\\left(\frac{N_r}{N^{1/2}} + \frac{N_\psi}{N^{3/2}} \frac{\partial N}{\partial \psi} \right) \label{sourcePoisson}\eea The problem here is akin to electrostatics, solving the equations subject to the boundary conditions that $\delta \vF_\psi$ vanishes (and $\delta \vF_r$ equate) at $\psi = 0$ and $\pi$, such that we avoid a jump in the field at $\psi = \pi/2$. 

\subsubsection{DM Regime}
For $r \ll r_0$, we can reduce (\ref{sourcePoisson}) and use an ansatz for $\phi$ of the form \be \phi = \frac{4\pi}{k} \frac{a_0}{\sqrt{r_0}}\,r^{3/2} \,F(\psi)\ee with profile function $F$: \bea  F \approx - 0.0236 - 0.1886 \, \cos 2\psi + 0.0108 \, \cos 4\psi \eea We can then compute the components of the MONDian force \bea -\Del\phi &=& \frac{4 \pi a_0}{\kappa} \left(\frac{r}{r_0}\right)^{0.5}
\left(\frac{3}{2}F \vect{e}_r + \frac{\partial F}{\partial \psi} \vect{e}_\psi\right) \eea \begin{figure} \begin{center}
\resizebox{1\columnwidth}{!}{\includegraphics{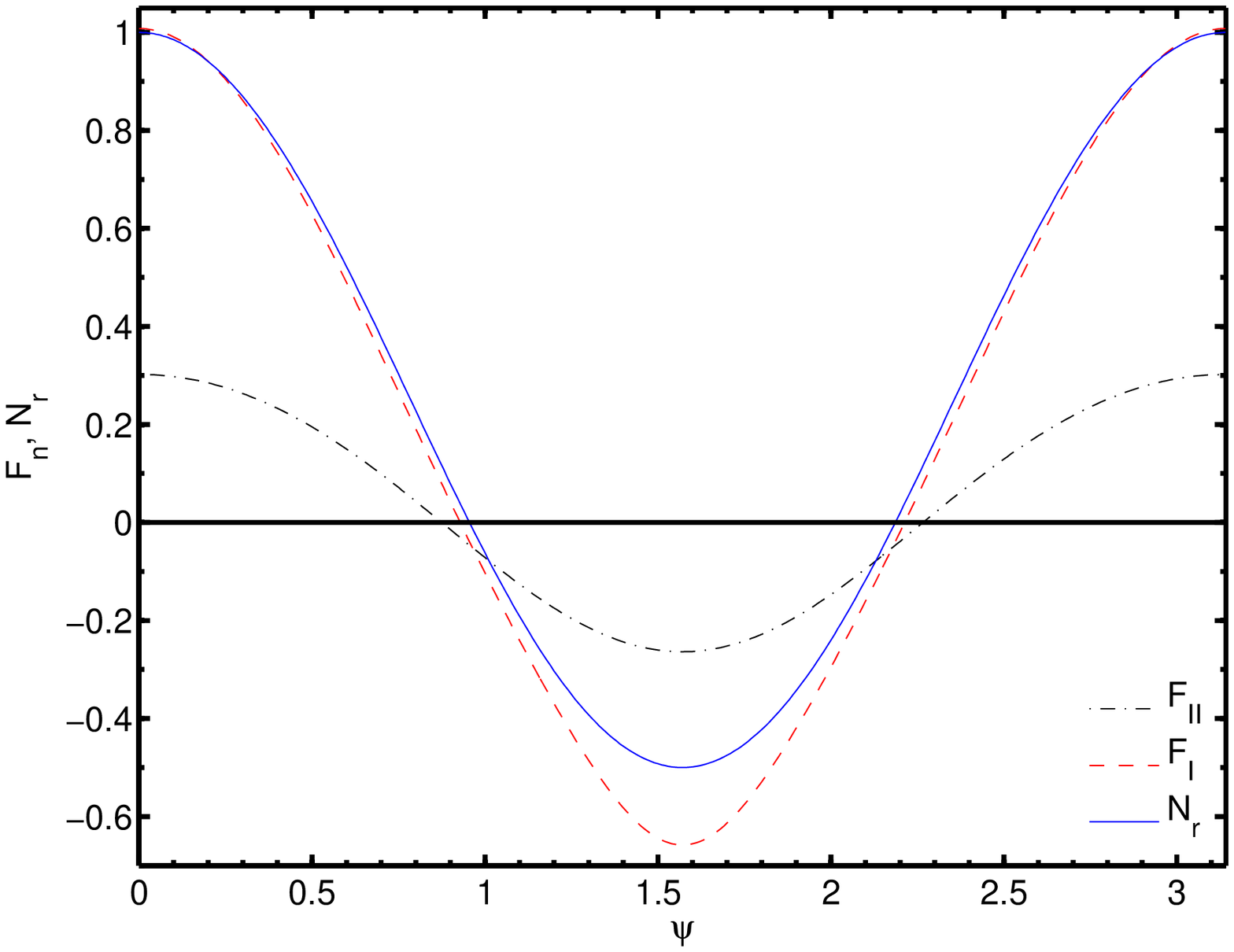}}
\resizebox{1\columnwidth}{!}{\includegraphics{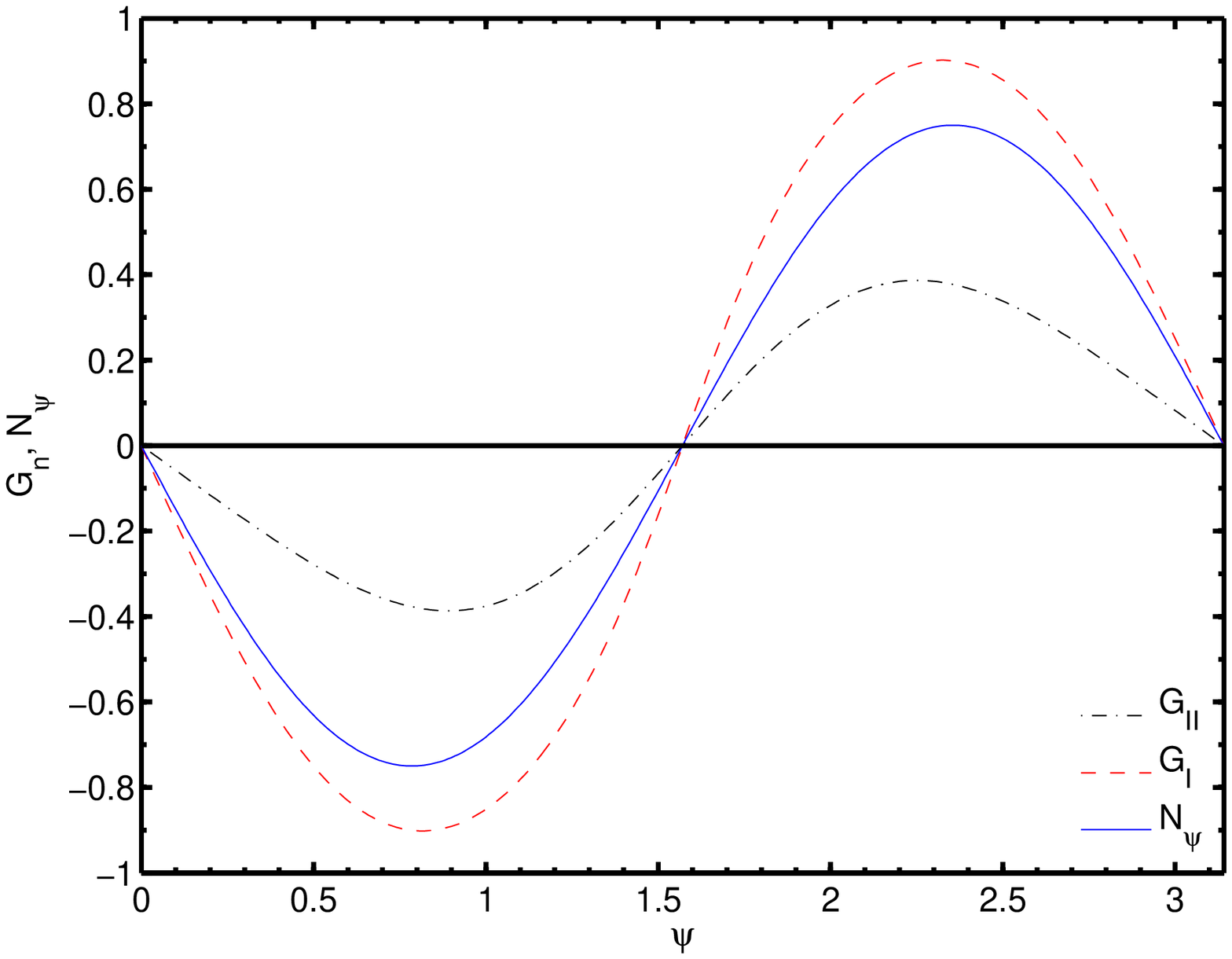}}
\caption{\label{fig:typeIIDMprofiles}{The angular profile functions F and G for the inner bubble forces in both type I and type IIB theories, alongside the linear Newtonian radial and azimuthal angular profiles. }}\end{center} \end{figure}and compare angular profile functions for type I and IIB solutions in Figure \ref{fig:typeIIDMprofiles}.  We see from the form of the force, \be \nonumber
\delta \vF = -\Del\phi = \frac{4 \pi a_0}{\kappa} \left(\frac{r}{r_0}\right)^p
\vect{S}(\psi) \Rightarrow S_{ij} \propto r^{p-1}\ee where in type I, $p \simeq 0.764$ and in type IIB, $p = 0.5$ - clearly the tidal stresses will
have a sharper divergence as we approach the SP.

\subsubsection{QN Regime}
For $r \gg r_0$, we similarly reduce (\ref{sourcePoisson}), pick an ansatz
to satisfy our boundaries conditions and then solve the resulting second
order ODE in $\psi$, \bea \phi &=& \frac{4 \pi a_0 r_0}{\kappa} \left(F(\psi) + \ln\left(\frac{r}{r_0}\right)\right) \\ F &\simeq& -0.2292 + 0.2876 \,\cos 2\psi - 0.1163 \, \cos 4\psi \eea and we also have the background rescaled Newtonian contribution \be \frac{\kappa}{4\pi}\Phi_{N} = -\frac{4\pi a_0}{\kappa}\frac{r^2}{8 r_0}(1+3\cos 2\psi)\ee which obviously is the dominant contribution in the $r/r_0 \gg 1$ limit.  

\subsection{Computational Techniques}\label{SOR}

\begin{figure}[h!]\begin{center}
\resizebox{1\columnwidth}{!}{\includegraphics{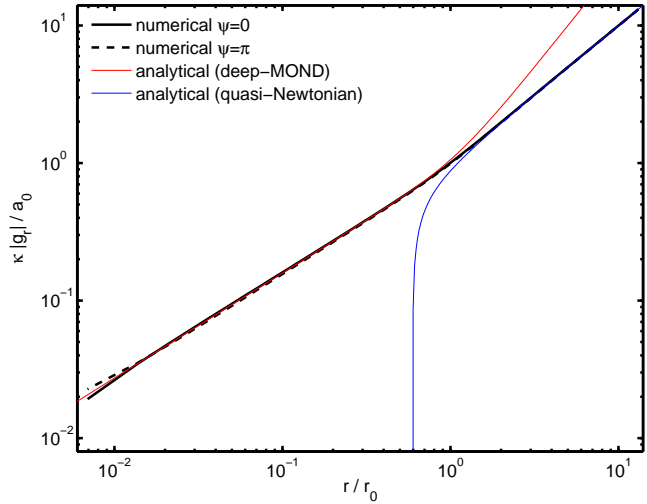}}
\caption{\label{fig:AnalyComp0}{A comparison between the numerical and analytical
results for components of $g = -\Del\phi$ for the Earth-Sun SP. Results are plotted as function of $r$ for $\psi = 0$ and $\pi$ (similar results are
found for other values of $\psi$, see Figure 2~\cite{bevis}).  As we see, the analytical symmetry between the $g$-component values (up to a sign) is also seen in the numerical case (except at low $r$, where the discretisation starts to become noticeable, giving us an estimate of the errors).  We used $C = 0.839$ for the DM scaling of the analytical results here.  Reproduced from data presented in~\cite{bevis}.}} \end{center}
\end{figure} 

\begin{figure}[h!]\begin{center}\resizebox{1\columnwidth}{!}{\includegraphics{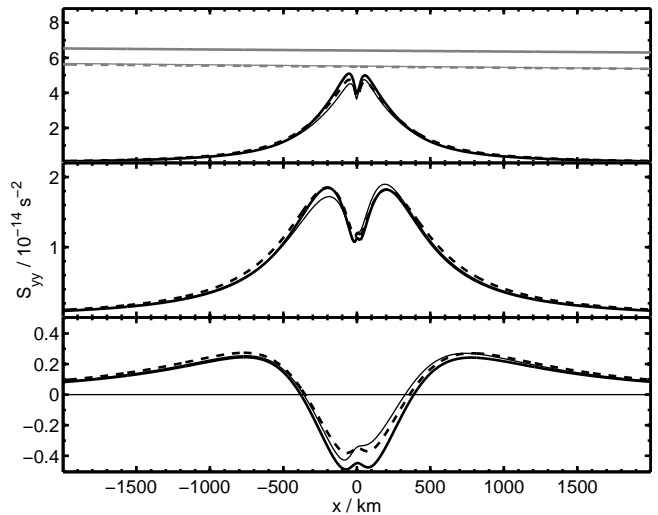}}
\caption{\label{fig:3bodyStress}The transverse MOND stress signal
$S_{yy}$ along $b=25$, $100$ and $400 \unit{km}$ (top to
bottom), for the Sun-Earth SP taking into account the effect of the Moon. The different lines represent lunar phases: new Moon (thick, black, solid), full Moon (thick, black, dashed) and the Moon appearing $18^\circ$ away from the Sun, towards positive $y$ (thin, black, solid). In the $b=25\unit{km}$ case, the Newtonian stresses (grey) rescaled by $\kappa/4\pi$ are shown for
comparison. Reproduced from data presented in~\cite{bevis}.}\end{center}
\end{figure}

Whilst the applicability of these analytical solutions is wide, they remain only strictly valid in the asymptotic regimes of large and small $r/r_0$.  In the intermediary regime (including the bubble boundary) very model dependent effects could be produced - suggesting we need a full numerical treatment of the system.  For the more complicated non-linear type I theories, we can
solve this system numerically, although the technical details are not important here (see~\cite{bevis}).  Suffice to say the codes used set up a non-uniform lattice around the SP with increased resolution for $r \ll r_0$ and attempts to simultaneously solve (\ref{vect1} - \ref{vect2}).  These results are then tested against the analytical solutions of Section \ref{Uformalism}, as seen
in Figure \ref{fig:AnalyComp0}.  We can use this code compute numerical solutions for this system, for both the Earth-Sun SP case, as well as a three body
Earth-Moon-Sun case.   Next we need to consider
the anomalous tidal stresses, which will need to take into account the rescaled Newtonian contribution from the MONDian field $\phi$:\be S_{ij} = -\frac{\partial^2 \phi}{\partial x_i \partial x_j} + \frac{\kappa}{4\pi} \frac{\partial^2 \Phi^N}{\partial x_i \partial x_j} \ee The results we find
from this tell us what the observable effect on measured tidal stresses in
the presence of a fifth-force field $\phi$.  We subtract off the rescaled
$\Phi_N$ contribution, which although plays the role of renormalising $G_N$
does not provide a real experimental observable.  A distinctive signal from
the $\vF_\phi \rightarrow S_{ij}^{\phi}$ should however provide a very good
observable.  Using our numerical
results, we plot the predicted tidal stresses along a given trajectory
past the SP in Figure \ref{fig:3bodyStress}, along with the introduction of the Moon in the dynamics.  As we see, the effect is only a perturbing one, the main dynamics still coming from the Earth and Sun.  

\section{A Route to Observables}
\subsection{LISA Pathfinder}\label{LISApathfinder}
LISA Pathfinder (LPF) presents the next generation of low frequency
gravitational wave interferometry instrumentation~\cite{LPF1, LPF2, LPF3}.  It is a technology validation mission for the Laser Interferometry Space Array (LISA) experiment~\cite{LISAsat}.  LISA's goal is to accurately detect  gravitational waves (GW) from astrophysical sources
using a space based laser interferometry.  Passing GWs induce oscillations
along the laser beams between the spacecraft (arranged in a triangle with
an inter-spacecraft distance of $5\times10^6$ km) and by monitoring these, we should be able to precisely measure GWs from, say, massive black hole
mergers and other extreme gravitational events.  The idea behind LPF is to emulate one of the arms of LISA by putting two test masses in gravitational free-fall, control and then measure their motion with unrivalled accuracy. In the process it will use and test a drag-free control system, a laser metrology system, inertial sensors and an ultra-precise micro-propulsion system.  Additionally the sensitivity of LPF is aimed at being more than two orders of magnitude better than any current experiment.  The nominal requirements of the mission are to: \begin{itemize}
\item{Test feasibility of laser interferometry with resolution approaching 10$^{-12}$ m Hz$^{-1/2}$ in the low frequency band of 1-30 mHz.}
\item{Demonstrate drag-free and attitude control in a spacecraft with two free proof masses.}
\item{Test the feasibility and endurance of the instruments in space.}
\end{itemize}

\begin{figure}[h!]\begin{center}
\resizebox{1\columnwidth}{!}{\includegraphics{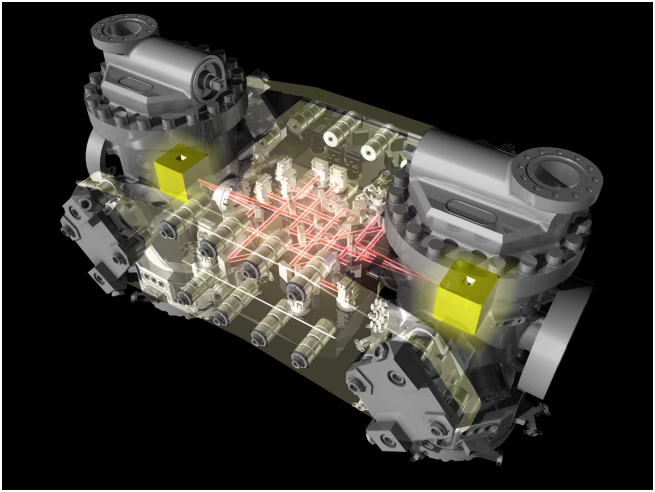}}
\caption{\label{fig:LPF}{The LTP demonstrating the laser interferometry between the two test masses. Reproduced from http://sci.esa.int/lisapf.}}
\end{center}\end{figure} 

The onboard instruments in the LISA technology package (LTP) are sensitive to the test mass motions relative to the spacecraft of up to 10$^{-9}$m and the relative test mass motion of up to 10$^{-12}$ m. The LTP will carry two identical proof masses, in the form of 46 mm cubes, made of gold-platinum each suspended in a vacuum can, as seen in Figure \ref{fig:LPF}.  The idea being to scale down an arm of LISA, from millions of kilometres to just centimetres.  The onboard disturbance reduction system includes a set of micro-rockets that aim to control the spacecraft's position to within 10$^{-9}$m. The drag-free control system consists of an inertial sensor, a proportional micro-propulsion system and a control system. The inertial sensors will monitor the micro motions of the two test masses and if they move away from their null positions, a signal is sent to the control system which is used to command the micro-propulsion thrusters, which in turn enable the spacecraft to remain centred on the test mass.

The LPF launch is planned for 2015, where it will head to the L1 Lagrange point with the operational phase lasting six months, with the possibility of extension up to one year.  After this, the remaining fuel could be used to make manoeuvres towards an Earth-Sun SP fly-by.  It may also be possible to include a second approach towards the SP and perhaps approaching the Moon-Earth-Sun saddle is an additional possibility.    

\subsection{Gravitational Wave Techniques}\label{GWtech}
Predictions are cast in the form of tidal stresses, because 
this is what will be measured by the instrument. LPF measures  
the relative acceleration between the masses (or its  Fourier transform (FT) in time) and up to a factor dependent on the proof mass separation, the
measurement is therefore one of tidal stress along the direction
linking the two masses (with further masses, other tidal stress
components would become accessible). In line with this statement,
noise evaluations and forecasts are expressed in terms of tidal stress
or relative accelerations; one should use the inter-mass separation (approximately
0.38m) to convert between the two.

The data analysis task in hand is therefore to detect a waveform of 
this type with the instrument aboard LPF. As a first hack at the problem, we evaluate the performance of noise matched filters. Matched
filtering is a well-known data analysis technique used for
efficiently digging a signal with a known shape out of  noisy
data~\cite{Helstrom,sathya}. The technique is extensively used in
the search for gravitational waves. The idea is to correlate a
time series $x(t)$ with an optimized template designed to provide
maximal signal to noise ratio (SNR), given the signal shape $h(t)$
and the noise properties of the instrument.  The signal $h(t)$ here, as measured
by LPF, will be the relative acceleration between the two test masses.  This
can then be converted into a tidal stress signal (as a function of $\vect{x}
= \vect{v} t$), although we will leave the exact details and form of the stress signal we want to characterise to Section \ref{secsnr}.  Generally we have $x(t) = h(t-t_a) + n(t)$, where $t_a$ is the signal ``arrival time''
and $n(t)$ is a noise realisation. We want to correlate $x(t)$ and
an optimal template $q(t)$, yet to be defined, according to: \be
c(\tau)=\int_{-\infty}^\infty x(t) q(t+\tau)dt\;  \ee where
$\tau$ is a lag parameter, giving us essential leverage if we
don't know $t_a$ a priori. The average of $c$ over noise realizations is
the expected signal, $S$, and its variance is the square of the 
noise in the correlator,
$N^2$; the forecast signal to noise ratio is therefore $\rho=S/N$.
A straightforward calculation (under general
assumptions, namely the Gaussianity of the noise, see~\cite{ali} for discussions
of relaxing this requirement) shows that $\rho$ is maximized by choosing a template
with Fourier transform: \be \label{opttemp}{\tilde
q}(f)=\int_{-\infty}^\infty q(t) e^{2\pi i ft} dt= \frac{{\tilde
h}(f) e^{2\pi i f (\tau - t_a)}} {S_h(f)} \ee and setting the lag
$\tau$ to the arrival time, $\tau=t_a$. Here $S_h(f)$ is the power
spectral density (PSD) of the noise, conventionally defined from
\be{\langle {\tilde n}(f){\tilde n}^\star (f')\rangle}=\frac{1}{2}
S_h(f) \delta(f-f')\ee The maximal SNR, realized by the optimal template, is then: \be
\label{SNR}\rho=\rho_{\mathrm{opt}} = 2 \left[ \int_{0}^\infty \,
df\frac{ \left |\tilde h(f)\right|^2 }{S_h(f)} \right]^{1/2}\ee Notice that the optimal template, $q(t)$, defined by (\ref{opttemp}) is a filtered version
of the signal $h(t)$, with a pass where the noise is low and a cut
where the noise in high. Additionally see that the optimal SNR given by
(\ref{SNR})  is not the energy in the signal but an integrated
signal power weighted down by the noise PSD.

These techniques are run of the mill in gravitational wave
detection, where the arrival time of a signal is often not
known.  A simple example being a chirping signal, even with a fair idea of the signals shape, we can't know when a binary coalescence is to take place. We therefore have to shift the template Fourier transforms, ${\tilde h}(f)$, by all possible phases, until the maximal SNR is obtained (should there indeed
be a signal). This adds an extra parameter to the fit and may also be the source of spurious detections. It affects the management
of $1/f$ noise and increases the false alarm rates (as effectively 
we have a number of trials equal to the total observation time 
divided by the duration of the template). This problem is absent in the context of our test, since we know where the saddle is and therefore where the
signal starts in the time-ordered series and so $t_a$ is known\footnote{Although
there will also be some intrinsic experimental variation in $t_a$, shifts
in it (even on the scale of km) should not produce large deviations in SNR
as the signals here are typically on the scale of $10^2$ km.}. A natural truncation in integration time $T$ is also present, simplifying $1/f$ dealings.

It has been estimated that the saddle can be pin pointed to about a kilometer and the spacecraft location determined to within 10 km even with even the most basic tracking methods - given that the computational grids have this sort of resolution, the effect on the SNRs compared to these should be negligible.  We should add that these uncertainties are of a practical, experimental nature rather than a theoretical one, it has been liberally estimated that the MOND saddle will not be shifted with respect to the Newtonian saddle by more than a meter.  Thus, we can simply set $t_a=0$ with an appropriate choice of conventions and set to zero the time lag $\tau$ in the correlator $c$, to achieve optimal results.  This
means that for all practical purposes, the starting time is indeed known and to the same degree of approximation so is the spacecraft trajectory and velocity with respect to the saddle.

\subsection{Characterising a test using SNR}\label{secsnr}

\begin{figure}[t!]\begin{center}
\resizebox{1\columnwidth}{!}{\includegraphics{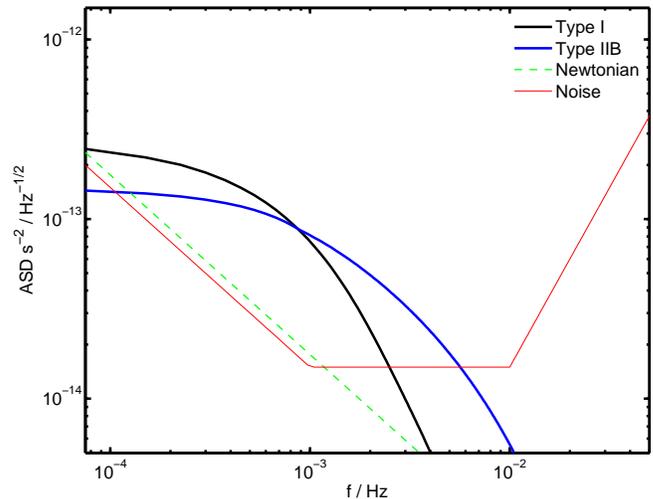}}
{\caption{\label{fig:signal + noise ASD}Here we plot the MONDian and rescaled
Newtonian signals, along with the noise profile, comparing type I and
IIB theories.  We have assumed a trajectory with the geometry described in the main text, with impact parameter of $b=50\;\unit{km}$ and velocity  $v=1.5\unit{km}\unit{s}^{-1}$ and have additionally plotted the contribution of $\phi$ to the Newtonian background.  This scenario produces SNRs of 28 and 35, for types I and IIB
(respectively).}} 
\end{center}
\end{figure}
The quantitative predictions for type I theories have been extensively studied using both analytical methods resorting to simplifying assumptions and numerical techniques~\cite{bevis}, including complications from the perturbing effect of the Moon and planets, as illustrated in Figure~\ref{fig:3bodyStress}.
 We adopt a coordinate system with $x$ aligned along the
Sun-Earth axis and centered at the saddle and considered 
trajectories parallel to $x$ ($y=b$ lines, where $b$ is the
impact parameter), but other trajectories are easy to implement.
Due to a number of practical issues~\cite{companion}, 
only transverse tidal stresses can be measured, 
say the $S_{yy}$ component. Recall the observable MONDian stress \be\nonumber S_{ij} = -\frac{\partial^2
\phi}{\partial x_i
\partial x_j} + \frac{\kappa}{4 \pi} \frac{\partial^2 \Phi^N}{\partial x_i \partial x_j} \ee remembering that the field $\phi$ produces both a MONDian effect and a rescaled Newtonian pattern, associated with a rescaling of $G$ in the Newtonian limit.  It is paramount that $\phi$ and $\Phi^N$ are found to the same degree of accuracy.  Given a spacecraft trajectory, the conversion of tidal stresses (such as those depicted in Figure ~\ref{fig:3bodyStress})
into a template in time, $h(t)$, is then trivial. For a setup such as the one described above we have \be h(t)=S_{yy}(vt,b,0)\ee where $v$ is the velocity of the spacecraft and $t=0$ corresponds to the point of closest saddle approach. In a more general setup, for an approximately constant velocity $\bf v$ with
a closest approach vector $\bf b$, and masses aligned along unit vector $\bf n$, we have: \be \label{hoft} h(t)=n^i n^j S_{ij}({\bf b} +
{\bf v}t )\ee This template should be Fourier transformed and using a given  noise model, used to produce an optimal template (using our noise matched filter techniques) so that finally its SNR can be evaluated.
\begin{figure}[h!]\begin{center}\resizebox{1\columnwidth}{!}{\includegraphics{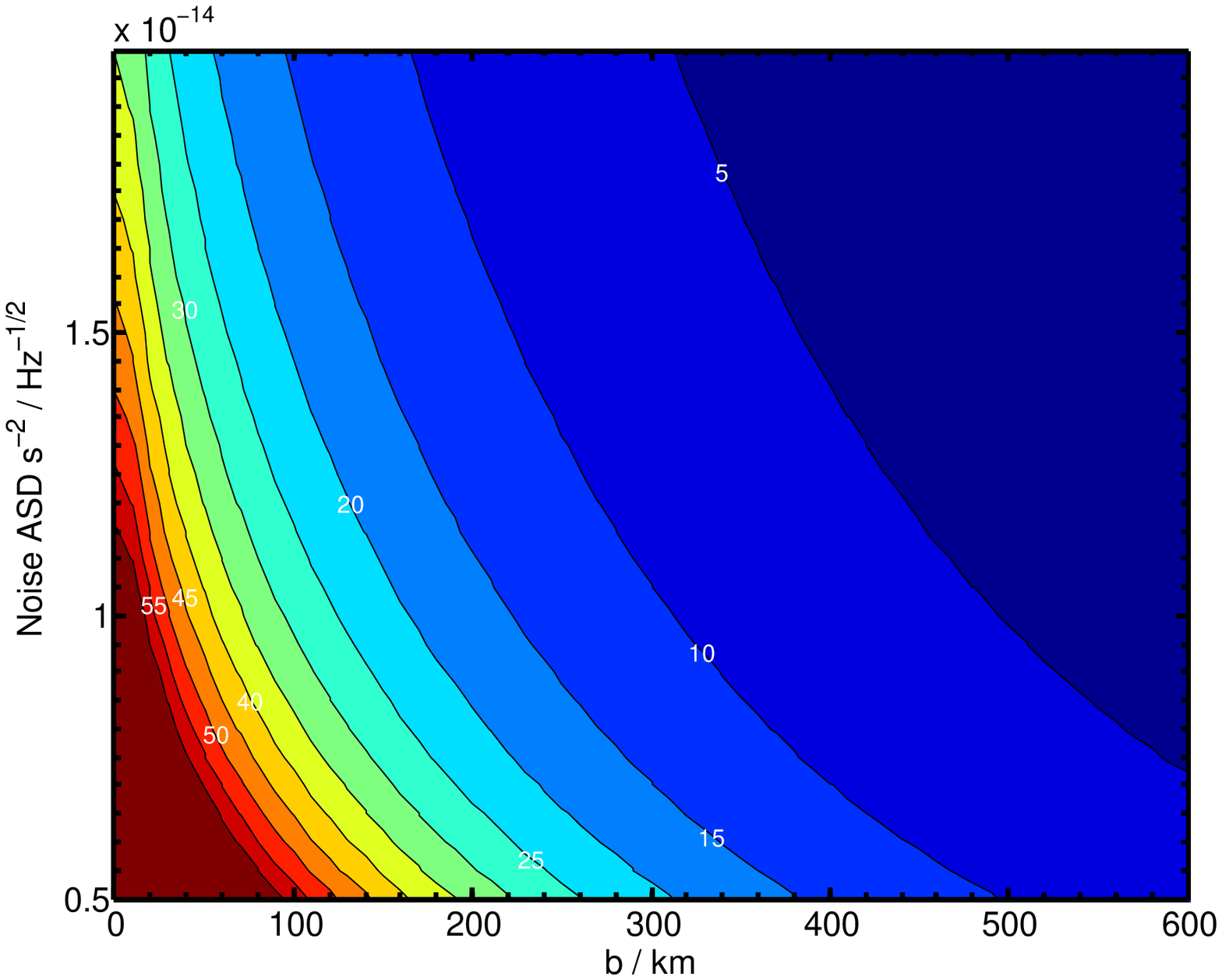}}
\resizebox{1\columnwidth}{!}{\includegraphics{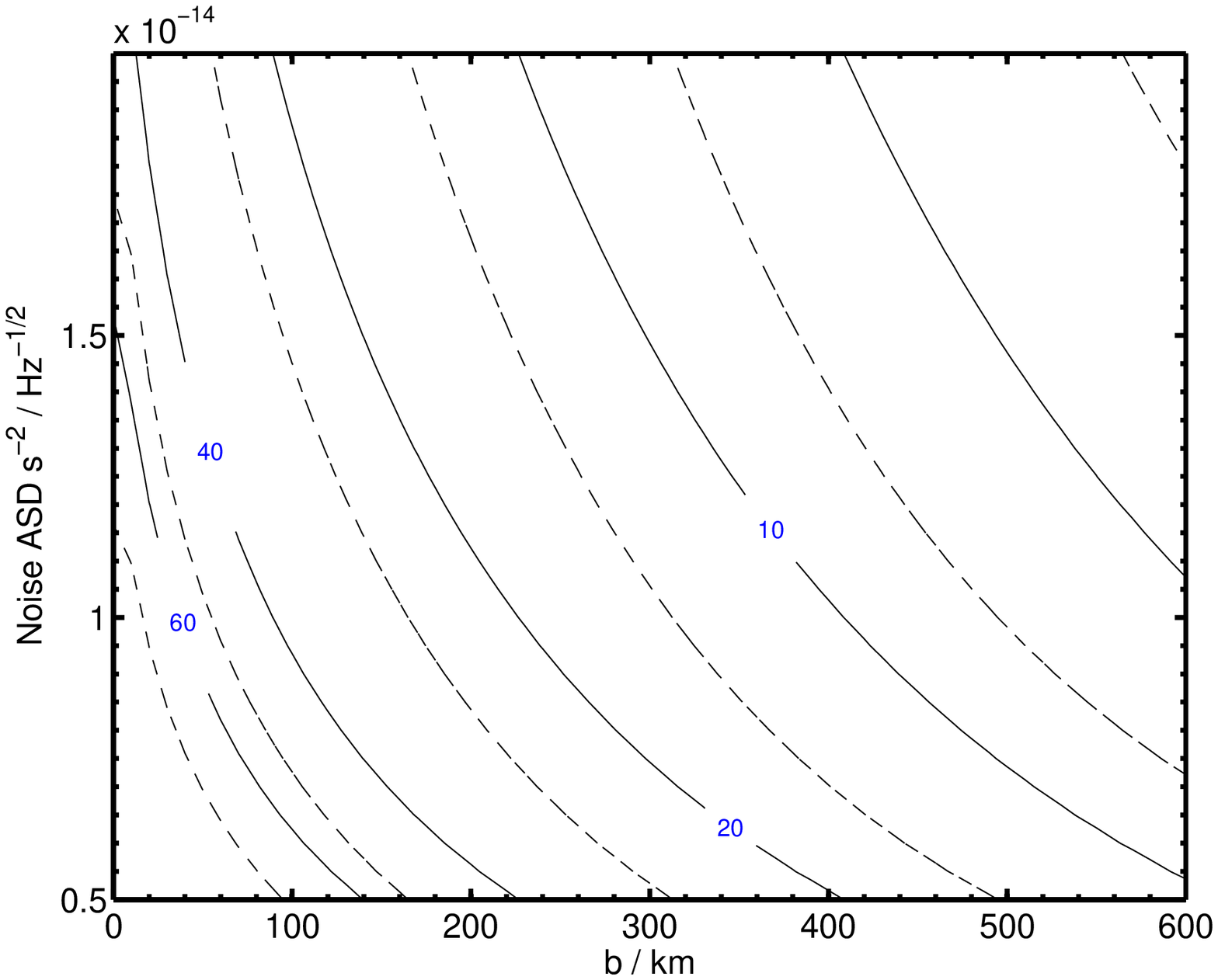}}
\caption{\label{fig:SNR contours} {\bf Top Panel} Signal to Noise ratio
contours, for various impact parameters up to 600km in type I theories, varying the idealised base noise ASD.  We set the spacecraft velocity at 1.5km s$^{-1}$. Calamitous assumptions would still lead to SNR of 5. More optimistic
ones ($b \sim 50$km with noise half way up the scale) would lead to
SNRs easily around 50. {\bf Bottom panel}, a comparison of SNR contour lines between type I and IIB theories.  The solid lines are the typical SNR to be obtained in IIB theories and the dashed lines to their immediate left the corresponding type I line - as we see IIB beats I.}\end{center} \end{figure}
We plot in Figure \ref{fig:signal + noise ASD} the amplitude spectral density (ASD) of the noise models as well as from the signal, \be P(f) = \frac{2}{T} \left| \int_{-T/2}^{+T/2} \diff
t \; h(t) \; e^{-2\pi i f t} \right|^{2} \ee where $f$ is the
frequency, $t$ is the time and $T$ is the integration period
(here taken conservatively to be $T= 2\times10^4\unit{s}$).  As a simplified
model (see~\cite{companion} for further details), we assume that the noise is white in the frequency range between $1$ and $10\unit{mHz}$, taking a constant baseline with ASD around $1.5\times10^{-14}\unit{s^{-2}/\sqrt{Hz}}$. For lower frequencies we assume $1/f$ noise and for higher frequencies that the noise degrades as $f^{2}$.  As we can see, there's signal to noise of order 10 over a couple of decades, making it not surprising that the integrated SNR is in double figures (in this case around 28).  Here we assume $v=1.5\unit{km} \unit{s}^{-1}$ and use impact parameters up to 600 km, varying the base line ASD of our noise model, as the results in Figure~\ref{fig:SNR contours} show
we would need to miss the saddle by more than 300 km to enter single figures in SNR for typical noise levels.  For $b \leq 50$ km, a SNR of 30-40 or greater
is not unrealistic (recent work has suggested $b \leq 10$ km is within easy
reach).

\begin{figure}[t!]\begin{center}
\resizebox{1\columnwidth}{!}{\includegraphics{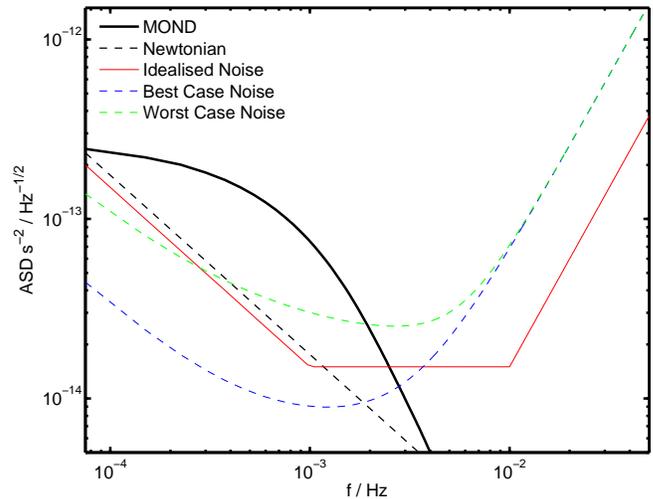}}
{\caption{\label{fig:realnoise}Here we replot Figure~\ref{fig:signal + noise ASD}, adding on the best and worst case scenarios for more realistic noise models (as at the time of writing). We have assumed a trajectory with the geometry described in the main text, with impact parameter of $b=50\;\unit{km}$ and velocity  $v=1.5\unit{km}\unit{s}^{-1}$ and have additionally plotted the contribution of $\phi$ to the Newtonian background.  These scenarios
generate SNR's of 12, 25, 37 for (respectively) the worst case, idealised
and best case noise.}} 
\end{center}
\end{figure}

 A number of improvements to the noise model are possible - one obviously being it is unlikely there will be a frequency region with white noise, instead it is likely to be higher than modeled at high frequencies but lower than expected at low frequencies. The turnover between the two regimes is smooth, as depicted in Figure~\ref{fig:realnoise}, where we superimposed the simplified noise model used before with the more realistic estimates for ASD for a best and worst case scenario.  This should not represent a major isue as these
forecasts will be run with the noise spectrum of LPF {\it in situ} at L1.

\subsection{The Earth-Moon Saddle}\label{moon}

\begin{figure}[t!]\begin{center}
\resizebox{1\columnwidth}{!}{\includegraphics{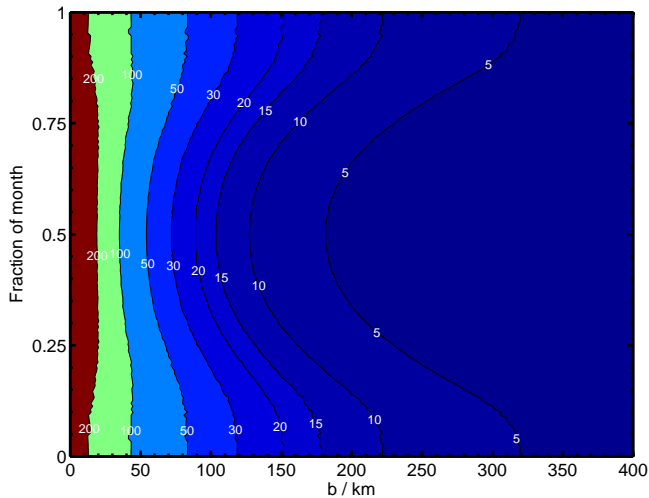}}
\caption{\label{fig:moon}{ SNRs for the Moon-Earth saddle, 
assuming our best case noise model and approach velocity $v=0.3\unit{km}\unit {s}^{-1}$, for different impact parameters 
and day of the month (0 and 1 represent the New Moon,
0.5 the Full Moon). We see that this saddle  is less forgiving if you miss it by more than 150 km and more rewarding if you get close to it (SNRs of 200 within reach).  If the former, we see that new moons generate higher SNRs.}}\end{center}
\end{figure}

Our techniques can also be applied to the issue of whether the Moon
saddle is a good alternative target. Practical matters may render
this saddle more amenable to multiple flybys, an issue that could be essential
in dismissing a ``false alarm'', should a positive detection be found.  As noted in Figure 10 of~\cite{bevis}, $r_0$ for the Moon saddle is smaller than the 380km found for the Earth-Sun saddle, and this size is more variable, depending strongly on the phase of the Moon (it varies between around 25km and 80km), however as $A$ is bigger, so the tidal stresses have a larger amplitude.  Nevertheless, what really matters for SNRs is the FT
of the signal as seen in time, with the satellite going through the bubble.
The large SNRs obtained for the Sun-Earth saddle result from a miraculous
coincidence between the sweet spot in the ASD,
and the size of the bubble as transformed into a time-signal by the 
typical velocities found in transfer orbits. This miracle could be spoiled
by the smaller size of the Moon saddle.  As it happens, orbits crossing the Moon saddle do so with a smaller velocity, typically smaller than $0.5\unit{km}\unit {s}^{-1}$. The two effects, smaller bubble combined with a lower speed, we find in fact counteract one another when converting the bubble signal into a time signal. Therefore it is not surprising that the SNRs predicted for the Moon saddle are as high as those for the Earth saddle (albeit more dependent on the phase of the moon).  

In Figure \ref{fig:moon} we plotted SNRs (assuming a Best Case noise model), for a crossing of the moon saddle at $v=0.3\unit{km}\unit {s}^{-1}$, with
varying impact parameters and for different days of the month. Here 0 and
1 on the $y$ represent the New Moon and 0.5 the Full Moon.  As we can see, in comparison with the Earth-Sun saddle, we find the Moon saddle:
\begin{itemize}
\item is less forgiving for $b > 150$ km.
\item is more rewarding if we here $b \leq 50$ km (with SNRs of perhaps even $\sim 200$ possible).
\item depends crucially on the lunar phase - New Moon produces the best results.
\end{itemize}
These results show there is great merit in including a moon saddle flyby into the considerations of LPF orbit designers (should a mission extension
occur).  

\section{Constraints from Data}\label{constrain}
\subsection{Negative Results}
\subsubsection{Constraining $(\kappa, a_0)$}

\begin{figure}[t!]\begin{center}
\resizebox{1\columnwidth}{!}{\includegraphics{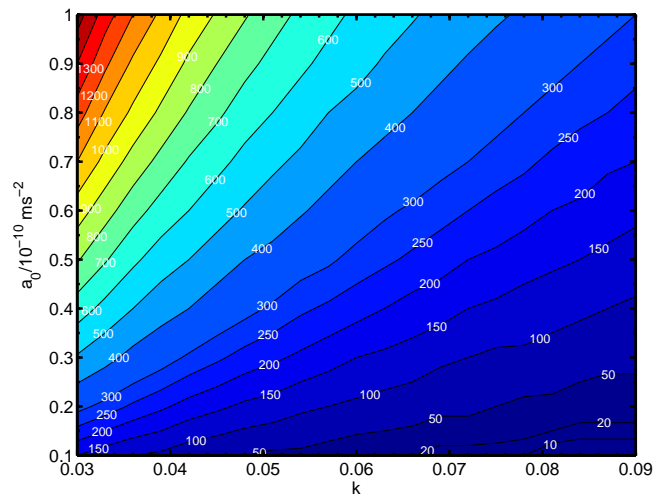}}
\caption{\label{fig:null}{Constraints placed on $a_0$ and $\kappa$ by a negative result for different impact parameters $b$ (labelling the lines and coding the colours).  We pick SNR $ = 1$ as the condition on our signal here and
then find an upper bound for $a_0, \kappa$ at a given impact parameter.  For a given $b$, the admissible parameter space would be ``outside'' the corresponding $b$ line (i.e. towards the right lower corner).}}\end{center}\end{figure}

Supposing we get a negative result, what constraints can we place upon $\kappa$
and $a_0$?  We may look for a preliminary estimate by seeking the region where the SNR for an optimal filter drops below 1, based on the bubble
size shrinking and thus reducing the prospects for signal.  This approach
is illustrated in Figure~\ref{fig:null} for various values of $b$ (where
$b$ labels the lines and codes the colours). For a given impact parameter, the admissible parameter space is ``outside'' the corresponding line (i.e. towards the right-bottom corner). In general, a negative result forces $a_0$ to be smaller and $\kappa$ to be larger than the fiducial values, the more so, the smaller the impact parameter $b$. As we see, if we were to miss the saddle by $\geq 1500$ km, the fiducial values of $\kappa$ and $a_0$ would survive a negative result. For an approach any closer, however, a negative result would rule them out and squeeze the parameter space towards the right-bottom corner. For $b\sim 10\unit{km}$, the $a_0$ (the $\kappa$) would have to be smaller (larger) than the fiducial values by an order of magnitude.  These constraints may now be combined with other pressures upon the theory, such as those arising from limits on $G$ renormalisation~\cite{Nconstraint}, Big Bang nucleosynthesis~\cite{CarrollLim}, fifth force Solar System tests~\cite{ssconst}, galaxy rotation curve data, and cosmological structure formation~\cite{tevesstrucform}.  However, as advocated in the introduction, by allowing complete freedom in $(\kappa,a_0)$, we can achieve a clear separation of the issues confronting these theories.

\subsubsection{Constraining Free Functions}
Putting aside detailed predictions for galaxy rotation curves
(which may well have been combined with inconsistent approximations, e.g.
regarding the curl field), the following criteria are reasonable
for what we will term physically permissible (or otherwise not fine-tuned) $\mu$ functions, defining type I theories:
\begin{itemize}
\item A. The cosmologically measured $G$ cannot differ significantly from that
measured, say, by the Cavendish experiment. That is: $G_{ren}\approx G$.
\item B. When the total Newtonian acceleration $a_N$ drops below $a_0$
the full potential $\Phi$ must be in the MONDian regime, that is, we need
$\phi$ to be in the MONDian regime {\it and} to dominate $\Phi_N$.
\item C. Function $\mu$ should only have one scale, below which $\phi$ is
MONDian, and above which it is  near Newtonian. Similar proposals can be
considered for $\nu$ functions in type IIB.  The detailed form of the transition is left undefined, but $\mu$ should have a single transition from $1$
\end{itemize}
\begin{figure}[t]\begin{center}
\resizebox{1\columnwidth}{!}{\includegraphics{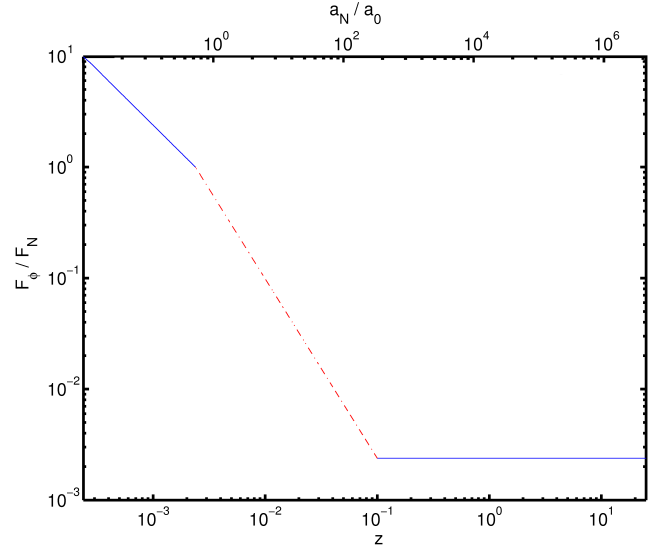}}
\caption{\label{fig:Fphi/FN plot null mu}{
Log plot of ratio between the MONDian and Newtonian forces, $F_\phi/F_N$, against $z=(k / 4\pi) |F_\phi| / a_0$ (bottom axis) and $F_N/a_0$
(top axis).  So that $F_N\sim F_\phi$ when 
$F_\phi\sim a_0$ (and so $z=\kappa/4\pi$; also $F_N\sim a_0$) 
and at the same time
have $F_\phi/F_N\sim
\kappa  /4\pi\ll 1$ in the Newtonian regime ($z\gg 1$, $F_N\rightarrow\infty$),
we must trigger MONDian
behaviour in $\phi$ at accelerations much larger than $a_0$.
However, by allowing a sharper intermediate power-law in $\mu$,
the trigger acceleration $a_N^{trig}$ may be smaller (in this illustration
by a factor of 10).} 
}\end{center}
\end{figure}

These considerations fully specify the function $\mu$, up to 
details on the transition regions.  Next consider the function:
\be
 \mu \rightarrow \left\{
                \begin{array}{llc}
                  z &&
 z<\frac{\kappa}{4\pi}\\ \\
\left(z/z^{trig}\right)^n
&&\frac{\kappa}{4\pi}<z<z^{trig}\\ \\
1& &z>z^{trig}
                \end{array}
              \right.
              \ee
such that the point where non-Newtonian behaviour in $\phi$ is triggered can be interchangeably pinpointed by:
\bea z^{trig}&=&\left(\frac{\kappa}{4\pi}\right)^{1-\frac{1}{n}}\\
a_\phi^{trig}&=&a_0\left(\frac{\kappa}{4\pi}\right)^{-\frac{1}{n}}\\
a_N^{trig}&=&a_0\left(\frac{\kappa}{4\pi}\right)^{-1-\frac{1}{n}}\eea
We note that $a_N^{trig}$ or $n$ can be swapped as a theory independent parameter, constrained by data. Clearly we are still insisting on requirement B, but also now with an intermediate region, where $\phi$ won't have dominated but is already non-Newtonian for $a_0<a_N<a_N^{trig}$.  As a result, the MOND bubble will shrink by \be r_0\approx 383 \left(\frac{\kappa}{4\pi}\right)^{\frac{n-1}{n}}
\unit{km} \ee From this, it is easy to see that changing $n$ closely from
2 will result easily in order of magnitude changes in $r_0$, but to reduce
it by more would require a some ``extreme'' intermediate power.

\begin{figure}[t]\begin{center}
\resizebox{1\columnwidth}{!}{\includegraphics{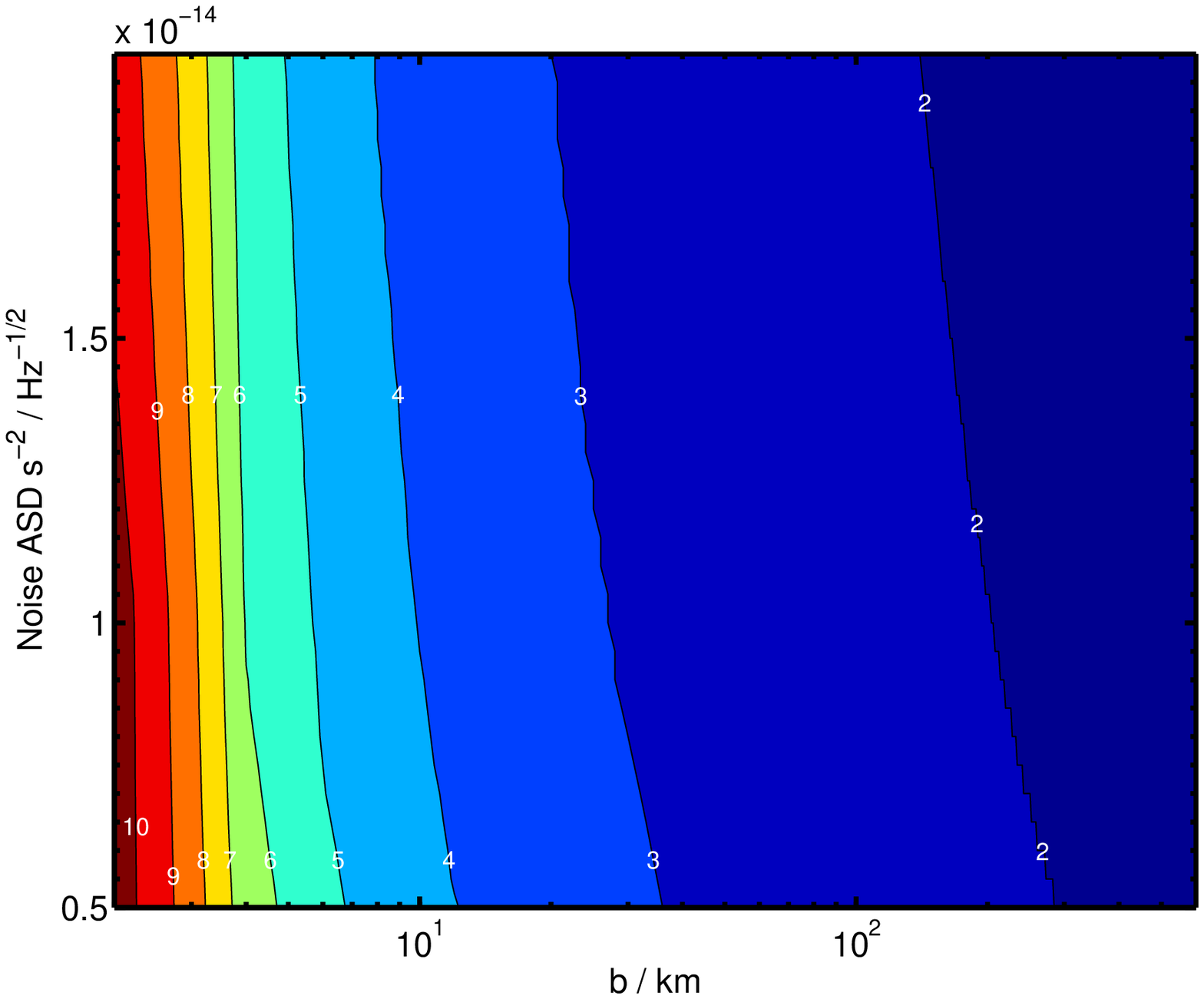}}
\resizebox{1\columnwidth}{!}{\includegraphics{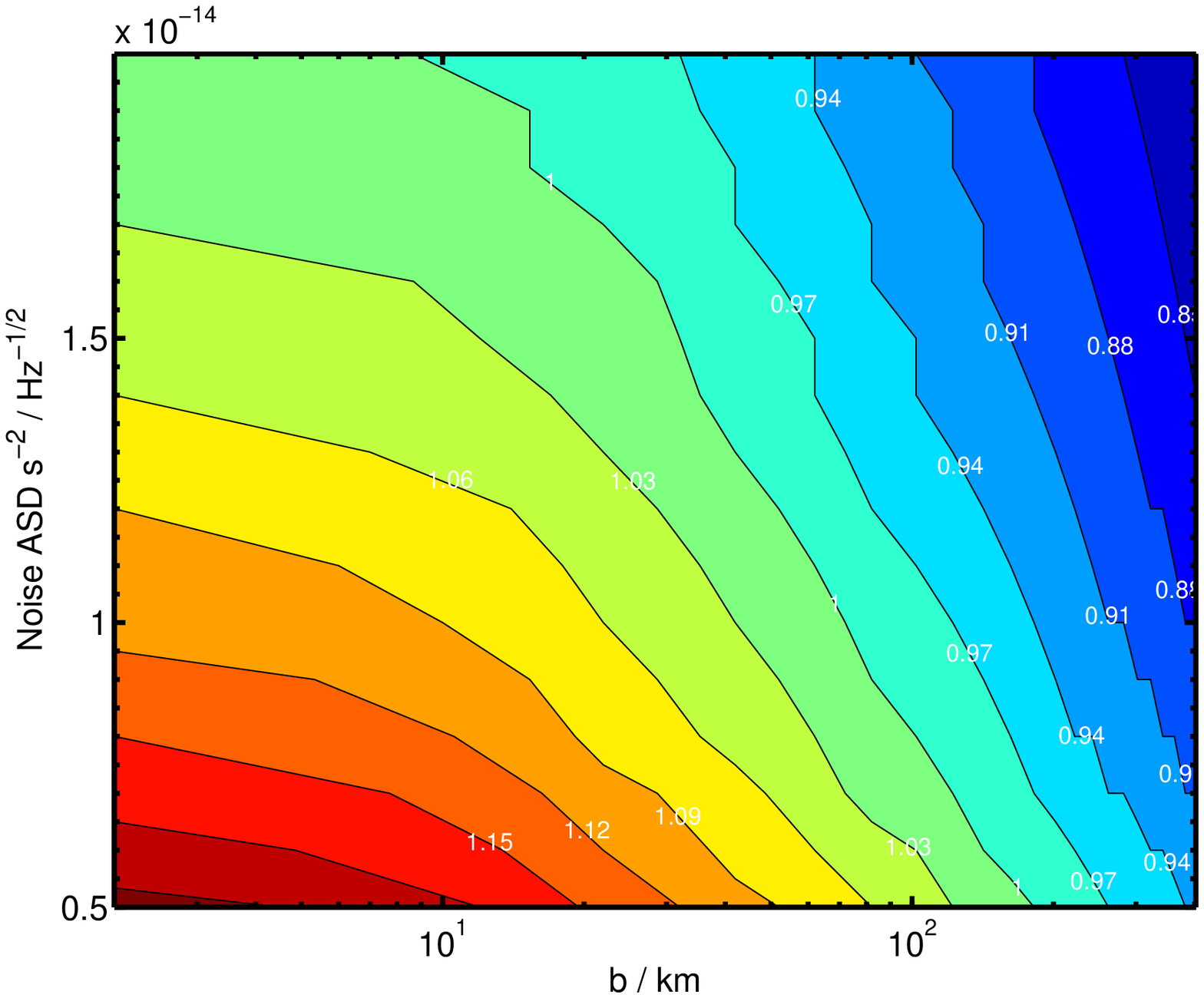}}
\caption{\label{fig:null SNR 1}{Contours of the power $n$ needed to
obtain SNR$=1$, with type I ({\bf top panel}) and type II theories ({\bf
bottom panel}).  Here
we use different noise levels and impact parameters
up to $b=400\unit{km}$, noting that these values of $n$ are an upper bound. For $n\neq 1$ the function is ``unnatural''. We see that as soon as we
plunge deep into the SP bubble, a rather unnatural ``designer'' free function
becomes necessary to accommodate a negative result.
}
}\end{center}
\end{figure}

Regrettably we can never make a model independent statement on what
$n$ is needed for a SNR of order 1. If nothing is observed, then 
by the nature of the problem, we must be making observations in the regime
$b\gg r_0(n)$.  Therefore we are necessarily probing the transient
from the far inner to the far outer bubble regimes.  Nonetheless it is interesting to perform an exercise of assuming a specific function with such behaviour, say:
\be \mu(z) = \frac{\left(z/z^{trig}\right)^n}{1 + \left(z/z^{trig}\right)^n}\ee
where $z_{trig}$ is considered to be smallest acceleration we would probe
whilst still seeing no observable anomalous behaviour.  
We can then proceed to fitting the values of $n$ which result in SNR = 1
values, as a way of seeing how extreme the functions have to be closer we
arrive at the saddle.  These results are condensed in Figure~\ref{fig:null SNR 1}, depicting the value of $n$ needed for a given $b$ and noise level.
 Obviously the value of $n$ produced here is merely an upper bound, from the condition on the SNR, larger values of $n$ would be acceptable too (although
clearly more ``unacceptable'').   As we can see as soon as we plunge deep into the MOND bubble, a rather unnatural designer $\mu$ becomes necessary to accommodate a negative result.

Similarly one can can proceed with a similar exercise with $\nu$ functions,
prescribing similar behaviour in the form of a function \be \nu = 1 + \left(\frac{w_{trig}}{w}\right)^n
\ee
the key difference being that the analogous result for the reduced bubble radius
is now \be r_0\approx 383 \left(\frac{\kappa}{4\pi}\right)^{\frac{2n-1}{n}} \unit{km} \label{tIIbubblesize}\ee and plot the results in Figure \ref{fig:null SNR 1}.  Thus our constraints between type I and IIB theories will be different and here the bubble size would be expected to shrink more, which given the lack of sharper divergence in the tidal stress makes sense.

\subsection{Positive Results}
From the scaling properties of our vacuum solutions~\cite{MagAliscaling}, we find that the anomalous tidal stresses must have the form \be S_{ij}=\kappa A H_{ij}{\left(\frac{\mathbf x}{r_0}\right)} \label{ts} \ee 
where the $H_{ij}$ is a universal (and complicated~\cite{aliscaling}) function, which hides the contribution of the free function.  This result has two interesting uses, firstly it allows us to generate templates for general values of $\kappa$ and $a_0$ from those for fiducial values, simply by rescaling them.  Secondly it allows us to make some progress at fixing the form of the free function in the event of a distinctive signal being found above the noise and Newtonian background.  
If we consider in type I theories, from the vacuum equations  \bea \Del\cdot\vU &=& 0 \nonumber \\ 4m\, U^2 \Del\wedge \vU + \vU\wedge \Del U^2 &=& 0\nonumber\eea and particularly the form of \be 4m = \frac{d \ln U^2}{d \ln \mu}\nonumber\ee we find: 
\begin{itemize} 
 \item{{ \bf DM regime}} For $z \ll 1$, $4m \rightarrow C^{DM}$, which becomes
 relevant when computing the exponent $\alpha(n)$ in the DM regime solutions.
 \item{{ \bf Departures from renormalised $G_N$}}, expanding $\mu$ in the $z \gg 1 $ limit gives \be \mu^{-1} \simeq 1 + \frac{C_1^\mu}{z^p} +\frac{C_2^\mu}{z^{2p}}+ \dots\ee  

\item{{ \bf QN regime}} For $z \gg 1$, $4m \rightarrow U^p / C^{QN}$ where $p$ is the leading order power relevant in the expansion of $m$, such that
we rewrite Equation (\ref{curlU2}) in the form of \be \Del \wedge \vU_2 = \underbrace{- \frac{\vU_0 \wedge \Del|\vU_0|^2}{\,|\vU_0|^{p+2} }}_{\Large{\Del \wedge \vU_2^r}} C^{QN}\ee where $\vU_2^r$ is the (renormalised) curl term (which
is just a function of $\vU_0$ and exponent $p$, but free of any scalings) and $C^{QN}$ is the model dependent scaling.  We find this is related according
to \be C^{QN} = \frac{p}{2}C^\mu_1\ee but we will
use this notation to be clear where each contribution arises from.  
\end{itemize}

Putting these together we find an expansion for $r \gg r_0$ of the form\bea \frac{\vU}{\mu} &\approx& \underbrace{\vU_0}_{\mathcal{O}(r^1)} + \underbrace{C_1^\mu \frac{\vU_0}{U_0^p} + \frac{p}{2}C^{\mu}_1\vU^r_2}_{\mathcal{O}(r^{1-p})} + \dots \eea where the higher order terms are $\mathcal{O}((r/r_0)^{1-2p})$ or smaller and represent more complicated combinations of $U_0$ and $U_2^r$.  For $b \gtrsim r_0$, the signal is sampling the QN regime, which means at lowest order \be \frac{\vU}{\mu} - \vU_0 \simeq \frac{\vect{h}(\psi)}{(r/r_0)^{p-1}}\ee
where we have collected together the angular function \be \vect{h}(\psi) = C^\mu_1 \left(\frac{\vN(\psi)}{N^p} + \frac{p}{2} \vB_p(\psi)\right) \ee Given that for small $n$, $h(\psi) \sim \mathcal{O}(1)$, we can first try fitting the radial fall off from the data (this should provide
the order of magnitude contribution) and then once the exponent $p$ is found, the various angular profile functions can be inferred, allowing us to get a bound on $C^\mu_1$.  In this way, we can start to reconstruct the form
of the $\mu$ function.  Similarly for $b \lesssim r_0$, we can attempt to fit the form of the tidal stress signal to proposed free functions.  An interesting
result found in the analysis of these theories~\cite{aliscaling,aliqumond} is that the inner bubble profiles do not change significantly for $\mu \rightarrow
z^a$ and $a \neq 1$ or $\nu \rightarrow 1/w^b$ and $b \neq 1/2$, meaning we are guaranteed a strong inner bubble signal
and that the best fitting for solutions will come from the outer bubble.  We can try to fit to parameterised free functions,
such as \bea \mu & = &\frac{z^a}{(1 + z^b)^{a/b}}\\ \nu &=& \left(\frac{1+ w^b}{w^b}\right)^{a/b}\eea  Such multi-parameter families of functions e.g.
$\mu(a,b)$ should be considered the ``minimal models'' that
we can use,
given that the two regimes here are constrained by complementary but different physical phenomena.  Our usual DM limit is motivated by this theory being a good dark matter replacement on low acceleration scales, but we consider dropping this requirement {\it a priori} and working with $\mu \propto z^a$. In the QN limit, the fall-off from $\mu \rightarrow 1$ is governed by agreement with bounds on fifth forces Solar System~\cite{ssconst}.  We see, therefore, it is prudent to consider at least a two parameter family of free functions.
 Analogous arguments exist for type II theories~\cite{aliqumond}, the only major difference being that the constraints will be different due to the
lack of curl field and this could be a decider in picking between models from
data.  


\section{Conclusions}\label{concs}

\begin{figure}[t!]\begin{center}
\resizebox{1\columnwidth}{!}{\includegraphics{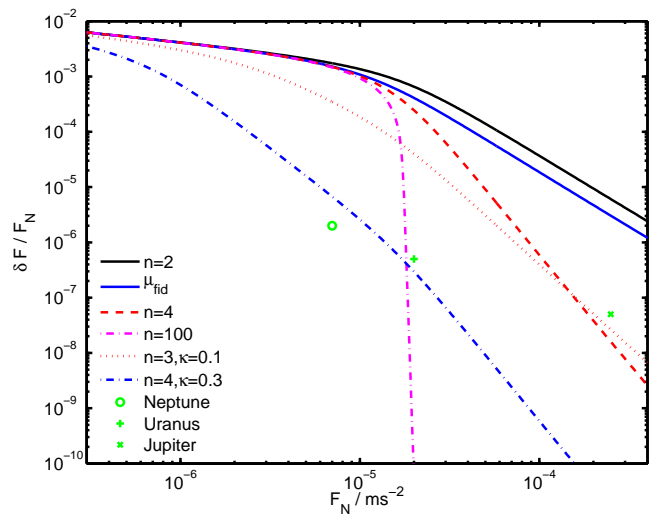}}
\caption{\label{fig:ssconstraints}{Comparing Solar System fifth force constraints~\cite{ssconst}
with models of free function.  Such a plot would a starting point to consider
current constraints on MG theories.  Recall that for fiducial parameter
values, the bubble boundary is at $a_{trig} \simeq \frac{4\pi}{\kappa}{a_0} \simeq 10^{-5} \unit{ms}^{-2}$ and here we have subtracted off the rescaled Newtonian contribution from the $\delta F$ and $\kappa = 0.03$ unless stated otherwise.  The function $\mu_{fid}$ corresponds to the $\mu$ of Equation \ref{mufid}.  The errors on the constraints from Uranus and Neptune remain high, even so note that our fiducial models would not satisfy the constraint from Jupiter - changes to either the fall off $n$ and/or $\kappa$ would be required. }}\end{center}
\end{figure}

In this work, we considered the case for testing and constraining theories of modified gravity with a preferred acceleration scale.  Such ideas were originally conceived in the guise of MOND as a replacement for dark matter but now have been elevated to fully relativistic, consistent alternatives to GR.  The weak field limits of these theories can produce phenomenology suitable for such purposes on galactic scales, however cosmological and other probes~\cite{tevescaustics, tevesstrucform, TimTom} can be problematic.  
In Section \ref{theory}, we showed how to attack these problems by first classifying the different modified poisson equations that result from these theories. Theories we labelled type I and IIB provide the best prospect for such a test, producing large regions of observable MONDian behaviour (due to the particular way their dynamics are triggered).  In such theories, the fifth force field $\phi$ present moves from taking the form of a rescaled Newtonian potential ($\phi \rightarrow \frac{\kappa}{4\pi} \Phi_N$, with $\kappa$ taken suitably small to escape detection on, say, Solar System scales) to producing a truly MONDian form.  When the total acceleration drops some preferred acceleration scale (which we denote $a_0$), this $\phi$ field becomes the dominant contribution.  Whilst this region around the Earth-Sun SP would only be $\sim 2.2$m in size, the anomalous tidal stress ``bubble'' of behaviour would be $\sim 383$km
 - providing a viable target for a satellite fly-by test.   We also considered
 the Earth-Moon SP and showing that provided we approach ``close enough'' and at the ``right phase'' in the Moon's cycle, very high SNRs are within arms reach, as illustrated in Figure \ref{fig:moon}.  The LPF space probe, designed to test the feasibility of space based, low frequency gravitational wave detection, could provide just the experimental test for a tidal stress experiment.  We find the peak of a potential MONDian tidal stress signal would be exactly around the lowest point in the expected noise spectrum - a useful coincidence.  Other theories, which we labelled types IIA and III, would fair less favourably, due to their one field approach for producing modified gravity effects - the effects which are the tiny saddle bubbles.  In Section \ref{secsnr}, we used the framework of experimental gravitational waves to estimate the SNRs for a LPF test.  Further to this, we considered experimental systematics such as different noise profiles,
spacecraft velocity and self gravity and as such the nominal requirements of the mission should be ample, as Figure \ref{fig:SNR contours}  shows.
 In Section \ref{constrain}, we considered how these Solar System tests would be able to constrain the parameter space, based on a null result.  Whilst a precise statement would be model dependent, we can obtain an order magnitude answer on the functional form of the free functions $\mu$ and $\nu$, as Figure \ref{fig:null SNR 1} shows.  Such constraints suggest it would be hard to wriggle out of a negative result (unless certain types of free functions are considered e.g. they diverge).  The different types of theory appear to have different behaviours in the regime we will be testing and so different constraints will apply to each - perhaps this can be used a discriminator between them.  We suggest therefore that a mission extension for LPF to probe these ideas would be scientifically feasible and provide good constraints on modified gravity theories (whatever the eventual result).

Looking to the future, we can consider a few prospects for additional work.  A consistent study of how to reconcile Solar System based constraints, such as the SP test and fifth force constraints, with galactic and other astrophysical settings for MOND should be made.  A proper fit of all the constraints available over {\it all} regimes has to date not been considered.  By way of a start, we can consider inner and outer Solar System constraints (e.g.~\cite{ssconst}) and see how our free functions compare on these scales, as outlined Figure \ref{fig:ssconstraints}.  A proper assessment of the weak-field limits of these theories should be considered, e.g. BiMOND~\cite{Milgrom:2009gv,Milgrom:2010cd} can produce different NR limits~\cite{qmond} using a different form of free function.  The FRW cosmology of such theories~\cite{TimTom} investigated generalisations of these theories and so the phenomenological implications of such should also be considered in this context.  It is well known preferred acceleration scale effects are not properly covered by the PPN formalism, e.g. time delay effects across MONDian bubbles as characterised in~\cite{MagTimeDelay}.  Perhaps a way forward would be developing a Post Parametrised Saddle formalism?  Characterising MOND from a more geometric point of view, as in~\cite{SkordisZlosnik} could be a starting point.

\section*{Acknowledgements}
The author would like to thank Jo\~ao Magueijo who first got them involved
in the hunt with modified theories, as well as Tom Zlosnik for useful comments.

\bibliographystyle{ieeetr}
\bibliography{references}


\end{document}